\documentclass[preprint,pre]{revtex4}
\usepackage{color,url,ulem,bm,amsmath,times,graphicx,epsfig}

\newcommand{\bmr}{\ensuremath{\bm{r}}}
\newcommand{\bmx}{\ensuremath{\bm{x}}}

\newcommand{\ei}{\ensuremath{\bm{e}_i}}
\newcommand{\eia}{\ensuremath{e_{i \alpha}}}
\newcommand{\eib}{\ensuremath{e_{i \beta}}}
\newcommand{\eide}{\ensuremath{e_{i \delta}}}
\newcommand{\eig}{\ensuremath{e_{i \gamma}}}
\newcommand{\bmu}{\ensuremath{\bm{u}}}

\newcommand{\cs}{\ensuremath{c_{\rm{s}}}}
\newcommand{\fieq}{\ensuremath{f_{i}^{\rm{eq}}}}
\newcommand{\fiseq}{\ensuremath{f_{i,s}^{\rm{eq}}}}
\newcommand{\myeq}[1]{Eq.\ (\ref{#1})}
\newcommand{\myfig}[1]{Fig.\ \ref{#1}}

\newcommand{\hide}[1]{{}}
\newcommand{\strike}[1]{{}}

\begin{document}
\title{Lattice Boltzmann study of pattern formation in reaction-diffusion systems}
\author{S.G. Ayodele$^1$, F. Varnik$^{1,2}$, and D. Raabe$^1$}
\affiliation{$^1$Max-Planck Institut f\"ur, Eisenforschung, 
Max-Planck Stra{\ss}e 1, 40237, D\"usseldorf, Germany.\\
$^2$Interdisciplinary Center for Advanced Materials Simulation,
Ruhr University Bochum, Stiepeler Stra{\ss}e 129,  44780 Bochum, Germany.}
\begin{abstract}
Pattern formation in reaction-diffusion systems is of great importance in surface micro-patterning [Grzybowski et al. Soft Matter. {\bf 1}, 114 (2005)], self-organization of cellular micro-organisms [Schulz et al. Annu. Rev. Microbiol. {\bf 55}, 105 (2001)] and in developmental biology [Barkai et al. FEBS Journal {\bf 276}, 1196 (2009)]. In this work, we apply the Lattice Boltzmann method (LBM) to study pattern formation in reaction-diffusion systems. As a first methodological step, we consider the case of a single species undergoing transformation reaction and diffusion. In this case, we perform a third-order Chapman-Enskog multiscale expansion and study the dependence of the Lattice Boltzmann truncation error on the diffusion coefficient and the reaction rate. These findings are in good agreement with numerical simulations. Furthermore, taking the Gray-Scott model as a prominent example, we provide evidence for the maturity of the LBM in studying pattern formation in non-linear reaction-diffusion systems. For this purpose, we perform  linear stability analysis of the Gray-Scott model and determine the relevant parameter range for pattern formation. Lattice Boltzmann simulations allow not only to test the validity of the linear stability phase diagram including Turing and Hopf instabilities, but also permit going beyond the linear stability regime, where large perturbations give rise to interesting dynamical behavior such as the so called self replicating spots. We also show that the length scale of the patterns may be tuned by rescaling all relevant diffusion coefficients in the system with the same factor while letting all the reaction constants unchanged.
\end{abstract}

\maketitle

\section{INTRODUCTION}
Spatially and/or temporally varying patterns have been observed in a variety of  physical~\cite{Arecchi1999,Wilson2008}, chemical~\cite{Epstein1996,De Wit1999,Sagu2003} and biological~\cite{Murray2002,Murray2003,Epstein2006,Kitsunezaki2006,Polezhaev2006,Schiff2007} systems operating far from equilibrium. The interest in understanding the  physics of pattern formation in these systems has been increasing steadily over the last few years especially after the experimental verification of Turing's idea~\cite{Agladze1992}. In chemical and biological systems for instance, macroscopic reaction-diffusion equations have been proposed as models for morphogenesis~\cite{Turing1952}, pattern formation~\cite{Murray2002,Murray2003} and self-organization~\cite{Nicolis1977,Tabony2002}. This class of equations usually includes the following two features: (i) a nonlinear reaction between chemical species describing local production or consumption of the species  and (ii) the diffusive transport  of these species due to density gradients. The simple form of the reaction-diffusion equation for a system of $N$ species is described by the following set of equations  
\begin{equation}
 \frac{\partial \rho_{s}(\bmx,t)}{\partial t}=D_{s}\Delta \rho_{s}(\bmx,t) + R_{s}, \quad 1\leq s \leq N,
 \label{eq:reac_diff1}
\end{equation}

where $\rho_{s}(\bmx,t)$ is the  mass density or concentration of species $s$ at time $t$ and location \bmx, $\Delta$ is the Laplacian operator with respect to spatial coordinate \bmx,  and $D_{s}$ is the diffusion coefficient of individual species $s$. In this work, we assume that $D_{s}$ is isotropic and independent of \bmx. The last term on the right hand side, $R_{s}$, is the reaction term. This term depends on the local density or concentration of the individual reacting species and the reaction mechanism governing the system. In most pattern forming systems, $R_{s}$  usually contains non-linear or autocatalytic reaction terms with product of the densities of the reacting species.

Due to their great importance both in biology, environmental science and industry, there has been growing interest in a study of these systems both experimentally, by numerical integration of the governing equations \text{and} via well-tuned analytic theories (see e.g.~\cite{Huang1993,Ruuth1995,Martin2002,Hundsdorfer2003,Kassam2005,Madzvamuse2006,Sebestikova2007} and references therein). However, solving problems with complex geometry (as is sometimes the case in biological systems) often requires a more efficient and robust method. The Lattice Boltzmann method has met significant success in simulating a wide range of phenomena in complex geometries over the last decades ~\cite{Varnik2007,Varnik2006,Succi1991,Rothman1990,Rothman1988}. In contrast with other traditional numerical techniques which only focus on the solution of the governing macroscopic equation, the Lattice Boltzmann method is based on kinetic theory. In cell-scale modeling of micro-organisms~\cite{Ishikawa2009,Pooley2008,Smith2006} for instance, the kinetic nature of the Lattice Boltzmann method makes the approach  computationally less demanding and allows for a relatively simple implementation of microbial interactions between cells. Furthermore, for problems involving large domain sizes, the local nature of LB operations allows easier implementation on parallel computational platforms thus enabling fast and large scale computations. In addition to the above features, the inherent capability of the LB approach in dealing with irregular boundaries, makes it suitable for studying reaction-diffusion phenomena in porous media~\cite{Kang2002} at the pore scale. However the accuracy and efficiency of a numerical method are often evaluated in terms of the smallest truncation error within the method. In previous Lattice Boltzmann studies of reaction-diffusion equation~\cite{Dawson1993,Blaak2000}, it is rather unclear as to how the truncation error varies with the system parameters such as reaction rate and diffusion constant. These parameters become important in pattern forming systems where non-linear reaction terms are present and reaction rate as well as diffusion constant may vary over a wide range.  Thus, for a better performance and accuracy, it is important to find out whether there is a range of optimal parameters that leads to the smallest truncation error and a better  convergence of the method. Such a study is performed in this work for the case of a single species reaction-diffusion systems. Performing a third-order Chapman-Enskog multiscale expansion, we investigate the dependence of the truncation error on the system parameters. Indeed, for this simple case, while the truncation error linearly varies with the reaction rate, it exhibits a pronounced minimum as a function of the diffusion coefficient. 

In order to extend the investigation to a pattern forming multi-species reaction-diffusion model, we have selected the Gray-Scott model~\cite{Pearson1993}, which serves as a standard paradigm for studying  reaction-diffusion systems. The Gray-Scott model, though simple, exhibits a wide range of interesting dynamical features including spots~\cite{Lee1994}, spiral waves~\cite{Muratov1999}, stationary waves~\cite{Doelman1997} and spatio-temporal chaos~\cite{Nishiura2001}. A particular feature of this model which makes it different from the other models is the existence of the so called self replicating spots~\cite{Reynolds1994}. Spatially localized cell like structures grow, deform and make replica of themselves. This act of ``cell division'' resembling DNA and RNA replication in cells or the replication growth of biological cells as seen in developmental biology makes it an ideal model  for  studying these biological systems with regard to pattern formation. In this reaction-diffusion system, generation of patterns comes usually from the instability of an initially uniform state to spatially inhomogeneous perturbations over a certain range of wavelengths. The possible range of wavelengths, as determined by a fixed set of system parameters, is usually invariant against a change of the system size. A change in system size often leads to a corresponding change in the number of spots, stripes or segments observed in the system. Hence, the number of segments or stripes is not invariant but proportional to the system size. In contrast, for some biological systems, the pattern forming wavelength is often proportional to the system size, while the number of stripes or segments is invariant against the change of system size. For instance, some mammalian coat markings have been shown to enlarge in proportion to system size~\cite{Kondo2002}, patterns in some micro organisms like \textit{Hydra} and \textit{Dictyostelium discoideum} have also been observed to show proportionality with size~\cite{Houchmandzadeh2002}. Modeling this type of biological systems with Turing-type reaction-diffusion therefore requires rendering the governing equations dimensionless and adjusting the system parameters in a proper way~\cite{Kaneko1994,Mizuguchi1995}. One such approach involves using diffusion constants which depend on the concentration of a system size-dependent auxiliary chemical factor  ~\cite{Othmer1980,Pate1984,Hunding1988,Aegerter-Wilmsen2005} or using the possibility that the concentration of some chemical changes with some power of the system size~\cite{Ishihara2006}. Interestingly, it is possible to change the length scale of the patterns in the Gray-Scott model via a simple rescaling of all the involved diffusion coefficients by the same factor, while keeping all the reaction constants unchanged. We provide a test of the validity of this simple approach with Lattice Boltzmann simulations.

The paper is organized as follows. In the following section, we briefly introduce the Lattice Boltzmann simulation scheme for reaction-diffusion equation. We then provide some benchmark tests for our LB simulation by comparing our results with  analytical solutions for the transformation reaction and diffusion of a point source in a domain with periodic boundary conditions. Excellent agreement with the analytical solutions is found. We also carry out a truncation error analysis of the model via a third-order multiscale expansion. Results obtained from this analysis are in agreement with our numerical simulations.  In  section \ref{sec:Gray-Scott}, we present the Gray-Scott model and, using linear stability analysis, determine  the parameter range for the existence of unstable solutions which we identify as a necessary condition for pattern formation. Our numerical simulations show good agreement with the predictions obtained from linear stability analysis. In section \ref{sec:scale_inv} we present a detailed study of the patterns which may be obtained via large amplitude perturbations of a linearly stable state. This case comprises the self replicating spots.

\section{The numerical model and its validation}

 \subsection{The lattice Boltzmann method}
\label{sec:method}
 The Lattice Boltzmann method~\cite{McNamara1988,Higuera1989,Qian1992,succi2001} can be regarded as a mesoscopic particle based numerical approach allowing to solve fluid-dynamical equations in a certain approximation, which (within, e.g. the so called diffusive scaling, i.e. by choosing $\Delta t=\Delta x^{2}$)  becomes exact as the grid resolution is progressively increased. The density of the fluid at each lattice site is accounted for by a one particle probability distribution $f_{i}(\bmx,t)$, where  $\bmx$ is the lattice site, $t$ is the time, and the subscript $i$ represents one of the finite velocity vectors $\ei$ at each lattice node. The number and direction of the velocities are chosen such that the resulting lattice is symmetric so as to easily reproduce the isotropy of the fluid~\cite{Rubinstein2008}.  During each time step, particles stream along  velocity vectors  $\bm{e_{i}}$ to the corresponding neighboring lattice site  and collide locally, conserving mass and momentum in the process. The LB equation describing propagation and collision of the particles is given by:
\begin{equation}
 f_{i}(\bmx+\ei, t+1)-f_{i}(\bmx, t)=\Omega_{i}f_{i}(\bmx, t),
\label{eq:fi}
\end{equation}
where $\Omega_{i}$ is the collision operator. 

The most widely used variant of LB  is the lattice BGK model \cite{Qian1992}, which approximates  the collision step by a single time relaxation towards a local equilibrium distribution $f^{eq}_{i}$. The lattice BGK model is written as:
\begin{equation}
 f_{i}(\bmx +\ei, t+1)-f_{i}(\bmx, t)=\frac{\fieq(\bmx, t)-f_{i}(\bmx, t)}{\tau_{\text{LB}}},
\label{eq:relax}
\end{equation}
where $\tau_{\text{LB}}$ is the relaxation time and the equilibrium distribution $\fieq$ is closely related to the low Mach number expansion of the Maxwell velocity distribution given as \cite{Frisch1987}
\begin{equation}
 \fieq(\bmx, t)= w_{i}\rho\left[ 1 + \frac{1}{\cs^{2}}(\ei \cdot \bmu) + \frac{1}{2\cs^4}(\ei \cdot \bmu)^{2}-\frac{1}{2\cs^{2}}u^{2}\right].
\label{eq:Maxwellian}
\end{equation}
In \myeq{eq:Maxwellian},  $\cs$ is the sound speed on the lattice ($c^{2}_{s} = \Delta x^{2}/3\Delta t^{2}$)  and $w_{i}$ is a set of weights normalized to unity. The weights $w_{i}$  depend on the number of velocities used for the lattice. In this work we use the two dimensional nine velocities (D2Q9) model with weights $w_{i}$ given as:
\begin{equation}
 w_{i} = \left\{ \begin{array}{ll}
         4/9 & \mbox{$\ei =(0,0),  i=0$};\\
         1/9 & \mbox{$\ei =(\pm1,0),(0,\pm1)\;\;  i=1....4$};\\
         1/36 & \mbox{$\ei =(\pm1,\pm1),  i=5....8$} \end{array} \right.
\label{eq:Weights}
\end{equation}

In order to model the reaction-diffusion equation in the frame work of a Lattice Boltzmann BGK model, we introduce a multi-species distribution function $f_{i,s}$ where the subscript $s$ runs over the number of species $s = 1...N$. In addition, as we are modeling  chemical reaction and  diffusion with no accompanying advection by the solvent  velocity, the mean flow velocity $\bmu$ in \myeq{eq:Maxwellian}  can be set to zero. This leads to

\begin{equation}
 \fieq(\bmr, t)=w_{i}\rho_{s}.
 \label{eq:equili}
\end{equation}
Equation (\ref{eq:equili}) satisfies the  requirement $\sum_{i=0}^N\fiseq= \rho_{s} $.
  The  chemical reaction is modeled  by including a source term, $R_{s}$, in the collision step. This leads to
\begin{equation}
 f_{i,s}(\bmx+\ei, t+1)-f_{i,s}(\bmx, t)=\frac{\fiseq(\bmx, t)-f_{i,s}(\bmx, t)}{\tau_{\text{LB},s}}+w_{i}R_{s},
\label{eq:relax}
\end{equation}

where $\tau_{\text{LB},s}$ is the relaxation time for species $s$. The source term $R_{s}$ represent the rate of change of density of the species, $s$, with regard to reaction kinetics. The exact form of the relation between the reaction rate  $R_{s}$ and the density (concentration) of each species depends on the type of reaction being modeled. The density  of the species $s$, $\rho_{s}$, is then computed from the distribution function using $\rho_{s}=\sum_{i=0}^Nf_{i,s}$.

Near equilibrium and in the limit of small Knudsen number (= mean free path/characteristic length of problem) the macroscopic  reaction-diffusion equation can be recovered using Chapman-Enskog multiscale analysis. The detailed analysis leading to the macroscopic equation is outlined in the appendix. The relaxation time $\tau_{\text{LB},s}$ is then found to be related to the diffusion coefficient as  $D_{s}=c^{2}_{s}\Delta t(\tau_{\text{LB},s}-\frac{1}{2})$.

\subsection{A transformation reaction}
\begin{figure}
\centering
\includegraphics[scale=0.60]{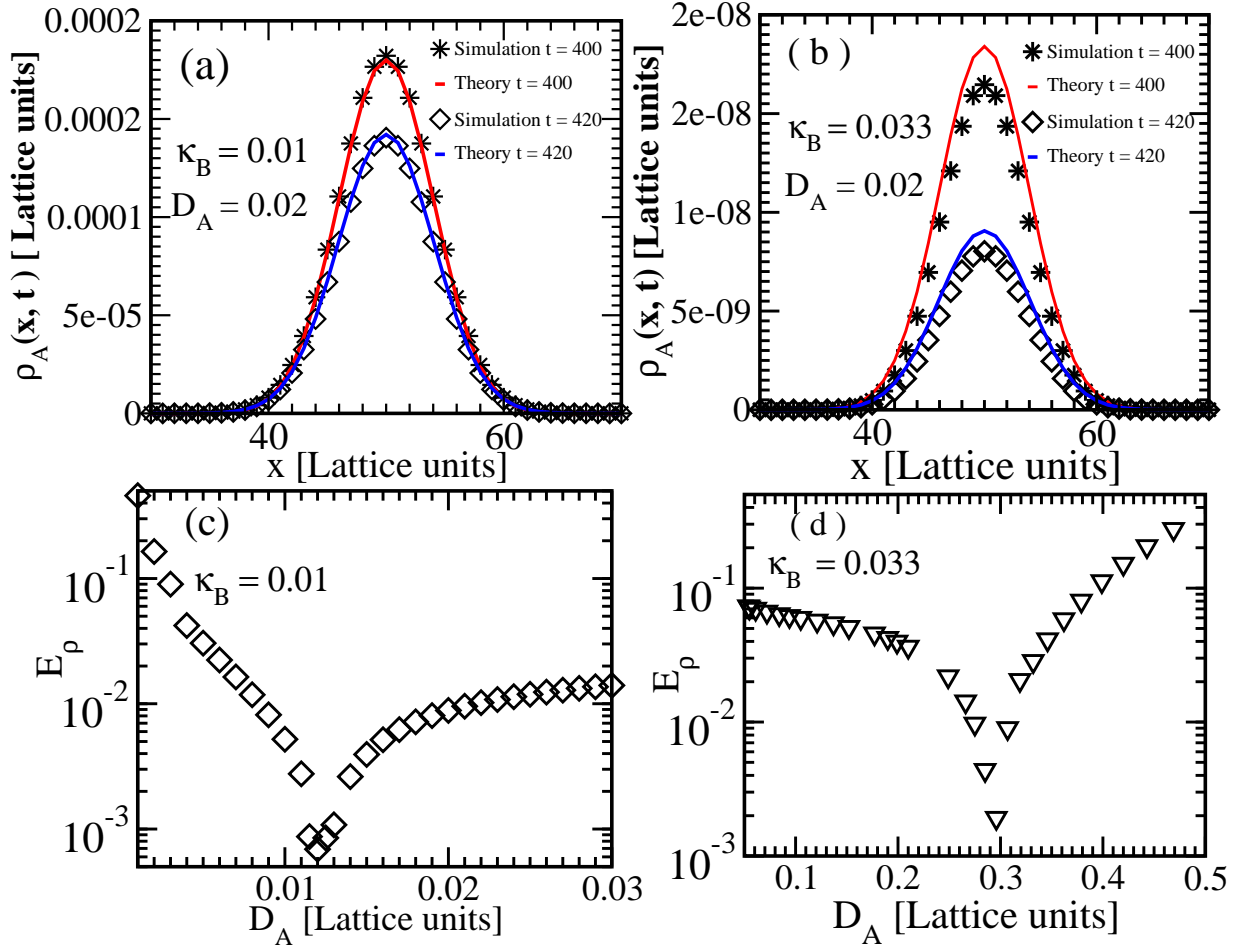}
\caption{ Comparison of the analytical solution in \myeq{eq:reac_diff5} and the Lattice Boltzmann simulation (a)    Density profiles of species A at t = 400 and t = 420 for $\kappa_{B}$ = 0.01 along the line $y= y_{0}$ (b) The same data as in (a) but for $\kappa_{B}$ = 0.033 (c) Behavior of the relative error in the density of A, $E_{\rho}$, with the diffusion coefficient $D_{A}$ at $\kappa_{B}$ = 0.01  (d) The same data as in (c) but for $\kappa_{B}$ = 0.033.}
\label{fig:graph1}
\end{figure}

 We provide here a simple test of our Lattice Boltzmann approach for  reaction-diffusion systems and characterize the truncation error obtained with regard to system parameters. We compute the problem of diffusion of a species A undergoing an irreversible transformation or decay reaction  to a species B,
\begin{equation}
A \stackrel{\kappa_{B}}{\longrightarrow}B.
\label{eq:reac_diff2}
\end{equation}
The reaction-diffusion equation describing the dynamics of species A can be written as   
\begin{equation}
\frac{\partial \rho_{A}(x,y,t)}{\partial t} = D_{A}\Delta \rho_{A}(x,y,t) -\kappa_{B} \rho_{A}(x,y,t), \quad 
\label{eq:reac_diff3}
\end{equation}
where $\rho_{A}(x,y,t)$ is the density of species A at point $(x,y)$ and time $t$, $D_{A}$ is the diffusion coefficient of A, $\kappa_{B}$ is the rate of the transformation reaction and $\Delta$ is the Laplacian operator with respect to spatial coordinates $x,y$. Using the initial condition $\rho_{A}(x_{0},y_{0},t=0) = \delta(x-x_{0})\delta(y-y_{0})$, Fourier transformation of \myeq{eq:reac_diff3} yields

\begin{equation}
\frac{d\hat{\rho}_{A}(q,t)}{dt} = D_{A}q^{2} \hat{\rho}_{A}(q,t) -\kappa_{B} \hat{\rho}_{A}(q,t), \quad  \hat{\rho}_{A}(q,0) = 1,
\label{eq:reac_diff4}
\end{equation}
 where $\hat{\rho}_{A}(q,t)$ is the Fourier transform of $\rho_{A}(q,t)$.   Integrating \myeq{eq:reac_diff4}, taking the inverse Fourier transform and slightly re-arranging the terms one obtains 
 \begin{equation}
\rho_{A}(x,y,t) = \frac{1}{(4\pi D_{A}t)} \exp\left( \frac{-(x-x_{0})^{2}-(y-y_{0})^{2}}{4D_{A}t}\right) \exp(-\kappa_{B}t).
\label{eq:reac_diff5}
\end{equation}
Equation (\ref{eq:reac_diff5}) is the analytical solution of the problem posed by \myeq{eq:reac_diff3} on a region infinitely extended in space. Setting $R_{A} = -\kappa_{B} \rho_{A}$ in the lattice BGK approach introduced in  \myeq{eq:relax}, we carried out numerical simulations for comparison with the above analytical solution. We set up a two dimensional domain of size $L_{x} =100$ and $L_{y}=100$ lattice units. At time $t = 0$, we set $\rho_{A}(x=L_{x}/2,y=L_{y}/2)=1$ , while $\rho_{A} =0$ on all other points  within the simulation box. For the whole region of the domain we initialize the density of species B to zero. We impose a periodic boundary condition in $x$ and $y$ directions and set the diffusion coefficient  of species A, $D_{A}= 0.02\Delta x^{2}/\Delta t$ and that of species B to $D_{B}= 0.02\Delta x^{2}/\Delta t$ . The results obtained are shown in \myfig{fig:graph1}. Note that at time of order $t_{D}\sim L^{2}_{x}/D_{A}$,  the effect of the boundary must be taken into account \cite{Ayodele2009}. Taking this into consideration, we have considered times of order $t< L^{2}_{x}/D_{A}$ in order to simplify our analysis. A comparison of the density profile of species A obtained at times $t=400$ and $t=420$ from the LB simulation and \myeq{eq:reac_diff5} is shown in \myfig{fig:graph1}(a) for $\kappa_{B}$ = 0.01. Quantitative comparison of the data is done by computing the relative error $E_{\rho}$ using the definition 

\begin{equation}
E_{\rho}=\sqrt{\frac{\sum_{x,y}{|\rho_{A,an}(x,y)-\rho_{A,sim}(x,y)|^{2}}}{\sum_{x,y}{|\rho_{A,an}(x,y)}|^{2}}},
\end{equation}

\begin{figure}
\centering
\includegraphics[scale=0.65]{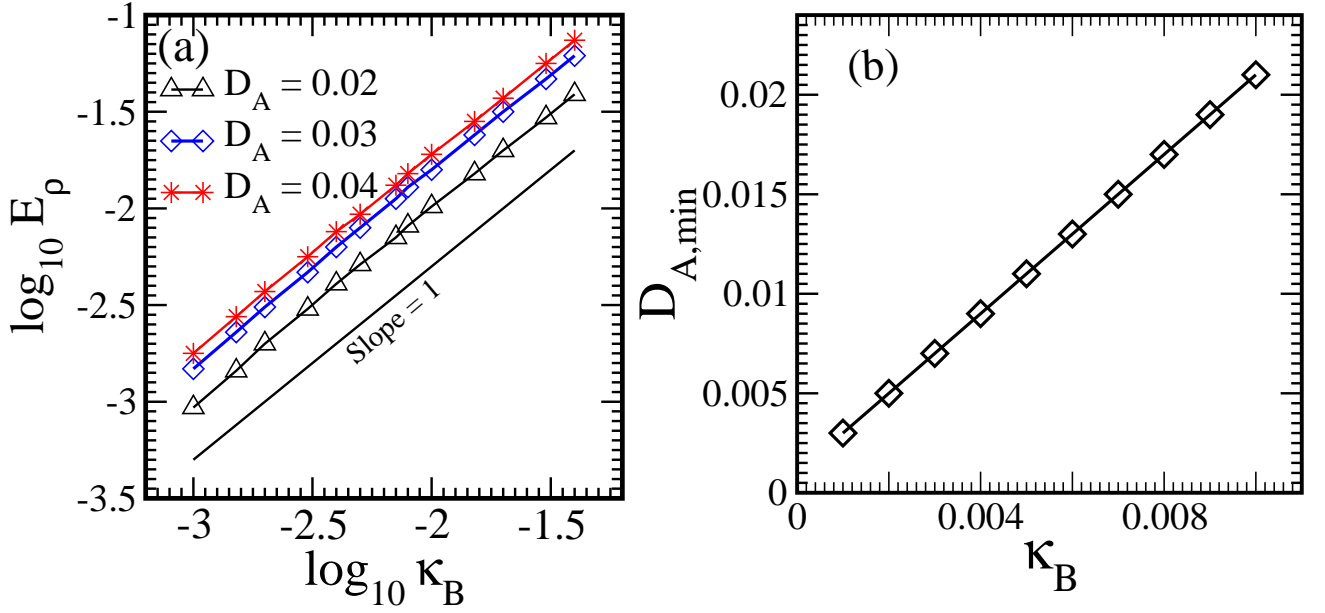}
\caption{(a) Log-log plot of $E_{\rho}$ versus $\kappa_{B}$ for different values of $D_{A}$. The curves are parallel to the solid black line with a slope of 1. (b) Plot of $D_{A,min}$ versus  $\kappa_{B}$. The curve is in line with \myeq{eq:DAmin}.}
\label{fig:graph2}
\end{figure}

where $\rho_{A,an}(x,y)$ is the density field obtained from the analytical solution in \myeq{eq:reac_diff5} and $\rho_{A,sim}(x,y)$ is the density field obtained from the simulation. The summation is taken over all lattice points in the domain.
We obtain a good agreement between the LB simulation and the analytical solution in \myeq{eq:reac_diff5}. The error in this case is less than 1\%. We note that increasing the value of the reaction rate $\kappa_{B}$ leads to an increase in the relative error as evident by comparing \myfig{fig:graph1} (a) for  $\kappa_{B}= 0.01$ and \myfig{fig:graph1} (b) for $\kappa_{B}= 0.033$. A heuristic explanation for this follows from the fact that an increase in $\kappa_{B}$ from 0.01 to 0.033 leads to a 3 times reduction in the time scale for reaction $t_{\rm R} = 1/\kappa_{B}$ . Thus, from a physical point of view, reactions occur three times more frequently within the same number of LB iterations. In other words, the relative length of the Lattice Boltzmann time step increases with regard to the characteristic reaction time, $t_{\rm R}$. As the LBM is linearly accurate in time, the error arising from the reaction part of LBM should exhibit a linear dependence on $\kappa_{B}$. As will be shown below this behavior is indeed substantiated by analytical error estimate and in our numerical simulations.

On the other hand, in the regime where $t_{R} \gg t_{D}$, one expects the effect of the reaction rate on the error to become less significant. Our numerical simulations indeed confirm this effect. To further explore this point, we systematically investigate  how the relative error in the concentration, $E_{\rho}$, varies with the diffusion coefficient $D_{A}$ and the reaction rate $\kappa_{B}$. Note that the computation of the error for each diffusion coefficient is done at the same \textit{physical} time corresponding to numerical times when the spatial extent of diffusion in all directions is the same. This ensures proper comparison across the range of different diffusion coefficients. The result of this investigation is shown in \myfig{fig:graph1} (c) and (d) for $\kappa_{B}= 0.01$  and  $\kappa_{B}= 0.033$ respectively. The first observation from both curves in \myfig{fig:graph1} (c) and (d) is the existence of a minimum in the magnitude of the relative error. It is interesting that a similar minimum in the LB truncation error is also observed in the computation of shear stress~\cite{Kruger2009}. Another  observation is the shift in the minimum of these curves. Figure \ref{fig:graph1} (c) shows that the minimum occurs at $D_{A,min}$ = 0.012, for  $\kappa_{B}= 0.01$,  while increasing $\kappa_{B}$ to 0.033 leads to a shift in  the minimum to $D_{A,min}$ = 0.03 (see \myfig{fig:graph1} (d)).  In order to validate the existence of this minimum in this model and characterize the shift observed for an increasing reaction rate, we perform a third-order Chapman-Enskog expansion of the Lattice Boltzmann BGK model for reaction-diffusion equation (see the appendix) and obtain an expression for the truncation error up to the third-order in the expansion parameter $\epsilon$. Using $R_{A}=-\kappa_B\rho_A$ in \myeq{eq:app23}, setting $\Delta t=1$ and rearranging the terms in orders of $\tau_A$ leads to
\begin{equation}
E =  3 c^{2}_{s} \partial_{t}\partial^{2}_{x_{\alpha}}\rho_{A} \left( \tau_{\text{LB},A}^{2}-(\frac{\kappa_B\partial_t\rho_{A}}{3 c^{2}_{s} \partial_{t}\partial^{2}_{x_{\alpha}}\rho_{A} } + 1 ) \tau_{\text{LB},A} + \frac{1}{6}\right).
\label{eq:Error}
\end{equation}
The third-order LB truncation error, $E$, is thus the product of time and spatial derivative of density with a quadratic polynomial in $\tau_A$. This polynomial has a minimum at 
\begin{equation}
\tau_{\text{LB},A, min}=\left(\frac{\kappa_B\partial_t\rho_{A}}{6 c^{2}_{s} \partial_{t}\partial^{2}_{x_{\alpha}}\rho_{A} } + 0.5 \right) 
\Rightarrow D_{A,min}=\frac{\kappa_B\partial_t\rho_{A}}{6 \partial_{t}\partial^{2}_{x_{\alpha}}\rho_{A}},
\label{eq:DAmin}
\end{equation}
where we also used the relation between diffusion coefficient and the LB relaxation time. Strictly speaking, the value of $D_{A,min}$ not only depends on the reaction rate $\kappa_B$ but also on space and time variables (through $\rho_A$). The fact that a minimum does indeed occur in $E_\rho$ as a function of $D_{A}$ is, therefore, not at all a trivial consequence of \myeq{eq:Error}. Indeed, the shape of $E_\rho(D_A)$ is not similar to a parabola suggesting that the non-trivial effects related to time and spatial derivatives of $\rho_A$ are present.

Nevertheless, it is worth testing to which extent useful information of the behavior of LB truncation error can be gained via the above analysis. For this purpose, we note two important features, which can be extracted from Eqs.~\ref{eq:Error} and ~\ref{eq:DAmin}. The first one is that $E_{\rho}$ could be a linear function of the reaction rate $\kappa_B$ (see \myeq{eq:Error}). The second observation is that also the position of the minimum in $E_\rho(D_A)$, i.e.\ the value of $D_{A,min}$ could be a linearly increasing function of $\kappa_B$. As illustrated in \myfig{fig:graph2}, results of Lattice Boltzmann simulations do confirm the validity of these two aspects. In summary, careful choice of the diffusion coefficient and reaction rate can lead to a better accuracy of the model. In a multi-species system with many reaction rate constants or wide range of time scales, the above discussion may provide guidance in choosing an optimal diffusion coefficient for each rate constant.

\section{The Gray-Scott model}
\label{sec:Gray-Scott}
We consider next the Gray-Scott model as a typical example of a two species reaction-diffusion system where the non-linear reaction terms between the species coupled with the transport by diffusion give rise to spatio-temporal patterns. The  Gray-Scott describes the kinetics of a simple autocatalytic reaction in an unstirred homogeneous flow reactor~\cite{Pearson1993}. The reactor is confined in a narrow space between two porous walls in contact with a reservoir.  Substance $A$ whose density is kept fixed at $A_{o}$ in the reservoir outside of the reactor is supplied through the walls  into the reactor with the volumetric flow rate per unit volume $k_{f}$. Inside the reactor, $A$ undergoes an autocatalytic reaction with an intermediate species $B$ at a rate $k_{1}$. The species $B$ then undergoes a decay reaction to an inert product $C$ at a rate $k_{2}$. The product $C$ and excess reactants $A$ and $B$ are then removed from the reactor at the same flow rate per unit volume $k_{f}$. The basic reaction steps are summarized as follows
\begin{equation}
     A + 2B  \stackrel{k_{1}}{\longrightarrow}3B,
\label{eq:auto_reac} 
\end{equation}
\begin{equation}
     B  \stackrel{k_{2}}{\longrightarrow}C.
\label{eq:decay_reac} 
\end{equation}
The reaction in \myeq{eq:auto_reac} is the cubic autocatalytic reaction in which two molecules of species B produce three molecules of B through interaction with the species A. The presence of B stimulates further production of itself, while the presence of A controls the production of B. Substance A is sometimes called the inhibitor and B the activator. By constantly feeding the reactor with a uniform flow of species A while at the same time removing the product and excess reactants, far from equilibrium conditions can be maintained. The equations of chemical kinetics which describe the above situations and include the spatio-temporal variations of the concentrations of A and B in the reactor take the following form:
       
 \begin{equation}
 \frac{\partial A}{\partial t}= k_{f} \left( A_{0}-A\right) -k_{1}B^{2}A+D_{A}\nabla^{2}A,
\label{eq:reac_diff10}
\end{equation}

\begin{equation}
 \frac{\partial B}{\partial t}= -\left( k_{0}+k_{2}\right)B +k_{1}B^{2}A+D_{B}\nabla^{2}B,\quad\quad
\label{eq:reac_diff11}
\end{equation}
where  $A$ and $B$ are the density of species A and B respectively, $A_{0}$ is the density of A in the reservoir, while $D_{A}$ and  $D_{B}$ are the diffusion coefficients of species A and B respectively. In order to understand and control the relevant time scale and length scale of patterns observed in the system, we introduce  variables in the form of time  and length scales that represent the physical processes acting in the system.  For A, the characteristic time scale is the time for the removal of A given as $1/k_{f}$ whereas for B it is $1/(k_{f}+k_{2})$. The characteristic time and length scales for these quantities are then:
\begin{equation}
  \tau_{A} = 1/k_{f},\quad\tau_{B} = 1/(k_{f}+k_{2}),\quad
   l_{A}=\left( D_{A}\tau_{A}\right) ^{1/2},\quad l_{B}=\left( D_{B}\tau_{B}\right) ^{1/2}.
  \label{eq:reac_diff12}
\end{equation}
Furthermore, we introduce the following dimensionless quantities:
\begin{equation}
 \tilde{A} = A/A_{0},\quad\quad \tilde{B} = B/B_{0},\quad\quad    
  B_{0}=\left( \frac{k_{f}}{k_{1}}\right) ^{1/2}.
  \label{eq:reac_diff13}
\end{equation}

  Introducing the quantities in Eqs.~(\ref{eq:reac_diff12}) and  (\ref{eq:reac_diff13}) into Eqs.~(\ref{eq:reac_diff10}) and  (\ref{eq:reac_diff11}) leads to:

\begin{equation}
 \tau_{A}\frac{\partial \tilde{A}}{\partial t}= -\tilde{B}^{2}\tilde{A}+1-\tilde{A}+l^{2}_{A}\nabla^{2}\tilde{A},
\label{eq:reac_diff14}
\end{equation}
and 
\begin{equation}
 \tau_{B}\frac{\partial \tilde{B}}{\partial t}= +\eta \tilde{B}^{2}   
\tilde{A}-\tilde{B}+l^{2}_{B}\nabla^{2}\tilde{B},
\label{eq:reac_diff15}
\end{equation}
where the parameter $\eta = \dfrac{A_{0}(k_{1}k_{f})^{1/2}}{(k_{f}+k_{2})}$ is the strength of the activation process. It adjusts the strength of the non-linear term in \myeq{eq:reac_diff15}.

The number of parameters can be further reduced by rescaling the time and length scale in units of $\tau_{A}$ and $l_{A}$ respectively. This  yields :
\begin{equation}
  \frac{\partial \tilde{A}}{\partial \tilde{t}}= -\tilde{B}^{2}\tilde{A}+1-\tilde{A}+\tilde{\nabla}^{2}\tilde{A},
\label{eq:reac_diff16}
\end{equation}

\begin{equation}
\frac{1}{\tau}\frac{\partial \tilde{B}}{\partial \tilde{t}}= +\eta \tilde{B}^{2}   
\tilde{A}-\tilde{B}+\frac{1}{\varepsilon^{2}}\tilde{\nabla}^{2}\tilde{B}.
\label{eq:reac_diff17}
\end{equation}
where $\tau=\tau_{A}/\tau_{B}$ and $\varepsilon=l_{A}/l_{B}$=$\sqrt{\tau_{A}D_{A}/\tau_{B}D_{B}}$. The parameter $\tau$  describes the relative strength of the reaction terms.

In general, equations (\ref{eq:reac_diff16}) and  (\ref{eq:reac_diff17}) are difficult to investigate by analytic means. However, simple cases exist for which analytical solutions can be found. We start with probably the most simple situation of a spatially homogeneous distribution of $\tilde{A}$ and $\tilde{B}$ ($\nabla^{2}\tilde{A}=0$,  $\nabla^{2}\tilde{B}=0$).  In this case, the steady state solutions  of  Eqs.~(\ref{eq:reac_diff16}) and  (\ref{eq:reac_diff17}) denoted as $\tilde{A_{e}}$ and $\tilde{B_{e}}$ obey 

\begin{multline}
\qquad\qquad\qquad\qquad-\tilde{B}_{e}^{2}\tilde{A}_{e}+1-\tilde{A}_{e} = 0 \text{,}\\
 \eta \tilde{B}_{e}^{2}\tilde{A}_{e}-\tilde{B}_{e} = 0.\qquad\qquad\qquad\qquad\qquad\qquad\qquad\qquad\qquad\quad
\label{eq:reac_diff18}
\end{multline}

Equation (\ref{eq:reac_diff18}) has three solutions. The first solution is the trivial homogeneous solution $\tilde{B_{e}}=0,\tilde{A_{e}} = 1$. This state exist for all system parameters.  The other two solutions exist provided  that $\eta > 2$. These are given by:
  
\begin{equation}
  \tilde{A^{\pm}_{e}} = \dfrac{\eta\pm\sqrt{\eta^{2}-4}}{2\eta},\qquad \text{and}\qquad \tilde{B^{\pm}_{e}} = \dfrac{\eta\mp\sqrt{\eta^{2}-4}}{2}.
\label{eq:reac_diff19}
\end{equation}

\subsection{Test of simulations in the case of spatially homogeneous dynamics}
\label{sec:LB_simulations_homogeneous}
\begin{figure}
\centering
\includegraphics[scale=0.60]{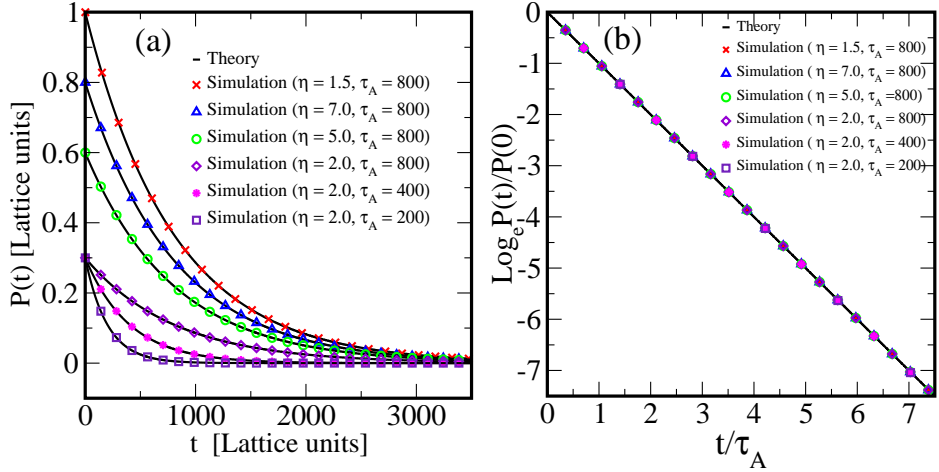}
\caption{(a)A plot of $-P(t)=\eta\tilde{A}(t)+\tilde{B}(t)-\eta$ versus time for $\tau=\tau_A/\tau_B=1$ but different values of the parameters $\tau_A$ and $\eta$ as indicated. The solid black curve in each case corresponds to theoretical prediction for the parameter values used in the simulation. In all studied cases, the simulation results exponentially decay to zero and show a perfect agreement with the analytical prediction, \myeq{eq:reac_diff21}. (b) Plot of $\text{Log}_{e}(P(t)/P(0))$ versus  $t/\tau_{A}$ for the same data as shown in (a).}
\label{fig:graph3}
\end{figure}

As a check of our simulation approach, we study the homogenized form of situations where Eqs.~(\ref{eq:reac_diff16}) and (\ref{eq:reac_diff17}) are accessible to an  analytical solution. For this purpose, we consider the case of spatially homogeneous dynamics with $\tau =1$, implying  $\tau_{A} =\tau_{B}$. In this case, multiplying \myeq{eq:reac_diff16} by $\eta$ and adding the result to \myeq{eq:reac_diff17} leads to

\begin{equation}
\frac{d (\eta \tilde{A} + \tilde{B})} {d\tilde{t}} = \eta - (\eta \tilde{A} +  \tilde{B}) \Rightarrow  \frac{dP}{d\tilde{t}} = -P
\label{eq:reac_diff20}
\end{equation}
 where $P = \eta  \tilde{A}+ \tilde{B} - \eta $ and we used the fact that $d\eta/d\tilde{t}=0$. Equation \ref{eq:reac_diff20} has the simple solution  $P(t)=P(0) \exp(-\tilde{t})=P(0)\exp(-t/\tau_A)$. In other words,
\begin{equation}
\eta \tilde{A}(t) + \tilde{B}(t) - \eta  =  [\eta \tilde{A}(0) -  \tilde{B}(0) - \eta ] \exp(-t/\tau_A)
\label{eq:reac_diff21}
\end{equation}
A test of \myeq{eq:reac_diff21} is provided in \myfig{fig:graph3} for  $\tau=1$ but different values of the parameters $\tau_A$ and $\eta$. In the case where $\eta < 2$,  the simulation starts from a spatially homogeneous state $\tilde{A}(0)=1$,  $\tilde{B}(0)=0$ with an additional density fluctuations  $\delta A =0.5$ and  $\delta B =0.25$ added homogeneously to A and B, respectively. This is done to break the symmetry which would keep the system at the initial state (due to the autocatalytic nature of the Gray-Scott model, without $B$, no reaction will take place). For all other cases where, $\eta \geq 2$ we start from the non-trivial states $(A^{\pm}_e,B^{\pm}_e)$  given by \myeq{eq:reac_diff19} with an additional small fluctuations of the form $\delta A =0.1$ and  $\delta B =0.1$. For all values of $\tau_A$ and $\eta$ investigated, a perfect agreement is found between theory and simulation.

\label{sec:spatially_homo}
\subsection{Stability analysis of spatially homogeneous states}
We proceed in this section to determine the stability of the stationary and homogeneous solutions obtained in  Eq.(\ref{eq:reac_diff19})  with regard to a spatially homogeneous perturbation. Our analysis starts by looking at the growth rate $\alpha$ of an infinitesimal perturbation about the steady state
\begin{equation}
\tilde{A} = \tilde{A_{e}}+\phi_{A}e^{\alpha t},\quad
\tilde{B} = \tilde{B_{e}}+\phi_{B}e^{\alpha t},
\label{eq:reac_diff22}
\end{equation}
where $\phi_{A}$ and $\phi_{B}$ are the amplitude of the perturbation to the species $A$ and $B$, respectively. Substituting \myeq{eq:reac_diff22} into Eqs.~(\ref{eq:reac_diff16}) and (\ref{eq:reac_diff17}), after linearizing and re-arrangement of the terms one arrives at the eigenvalue equation
\begin{equation}
   \left( \bm{J}-\alpha\bm{I}\right)\bm{\phi} = 0, 
\label{eq:reac_diff23}
\end{equation}
where $\bm{I}$ is the identity matrix, $\bm{\phi} = (\phi_{A},\phi_{B})^{T}$ and the matrix $\bm{J}$ is given as
 \begin{equation} 
 \bm{J}= \begin{bmatrix}
 \tau(2\eta\tilde{A}^{\pm}_{e}\tilde{B}^{\pm}_{e}-1)&\tau\eta\tilde{B}^{\pm^{2}}_{e}  \\\\
 -2\tilde{A}^{\pm}_{e}\tilde{B}^{\pm}_{e}& -(\tilde{B}^{\pm^{2}}_{e}+1)
  \end{bmatrix}.
\label{eq:reac_diff24}
\end{equation}
The eigenvalue equation in (\ref{eq:reac_diff23}) has the characteristic polynomial
\begin{equation}
  \alpha^{2}-\alpha\text{tr}\bm{J} +|\bm{J}| = 0,
\label{eq:reac_diff25}
\end{equation} 
 where $\text{tr}\bm{J}$ and $|\bm{J}|$ are the trace and determinant of matrix $\bm{ J}$. 
 The pair of solutions  or eigenvalues of matrix $\bm{J}$ is written

\begin{equation}
  \alpha_{1,2} = \frac{1}{2}\left( \text{tr}\bm{J}\pm\sqrt{(\text{tr}\bm{J})^{2}-4|\bm{J}|}\right).
\label{eq:reac_diff26}
 \end{equation}

  The eigenvalues $\alpha_{1,2}$ in \myeq{eq:reac_diff26},  can either be real or complex conjugate depending on the relative magnitude and sign of the determinant $|\bm{J}|$ and trace $\text{tr}\bm{J}$. If the real part of at least one eigenvalue is positive, the considered solution  is unstable. 

For the trivial state $(A_{e} =1,B_{e}=0)$, $\text{tr}\bm{J} = (\tau +1)$ and $|\bm{J}|=\tau$. Using \myeq{eq:reac_diff26} and the fact that $\tau > 1$ one obtains that both eigenvalues are negative. Hence this state is linearly stable with respect to spatially homogeneous perturbations. Next we consider the non-trivial stationary homogeneous solutions. In this case, inserting the solutions of $A^{\pm}_{e}$ and $B^{\pm}_{e}$ given in \myeq{eq:reac_diff19}  in \myeq{eq:reac_diff24}, one obtains that  $\text{tr}\bm{J}$ = $\left(\tau-\eta \tilde{B}^{\pm}_{e}\right) $ and $|\bm{J}|= \tau\left( \eta \tilde{B^{\pm}_{e}}-2\right)$. Furthermore, since in this case $\eta>2$, and using \myeq{eq:reac_diff19}, one can easily verify that $|\bm{J}|(\tilde{B}^{+}_{e})=(\eta\tilde{B}^{+}_{e}-2)<0$ and  $|\bm{J}|(\tilde{B}^{-}_{e})=(\eta\tilde{B}^{-}_{e}-2)>0$. Now from \myeq{eq:reac_diff26} it follows that, independent of the sign of $\text{tr}\bm{J}$, one of the solutions $\alpha_{1,2}$ is always positive, provided that $|\bm{J}|<0$. Hence, the state $\tilde{B}^{+}_{e}$ is always unstable. For the state $\tilde{B}^{-}_{e}$, on the other hand, both solutions $\alpha_{1,2}$ will have the same sign as $\text{tr}\bm{J}$ and thus the state $\tilde{B}^{-}_{e}$ may be stable provided $\text{tr}\bm{J} < 0$  (i.e.\ $\tau< \eta\tilde{B}^-_e$).

\begin{figure}
\centering
\includegraphics[scale=0.60]{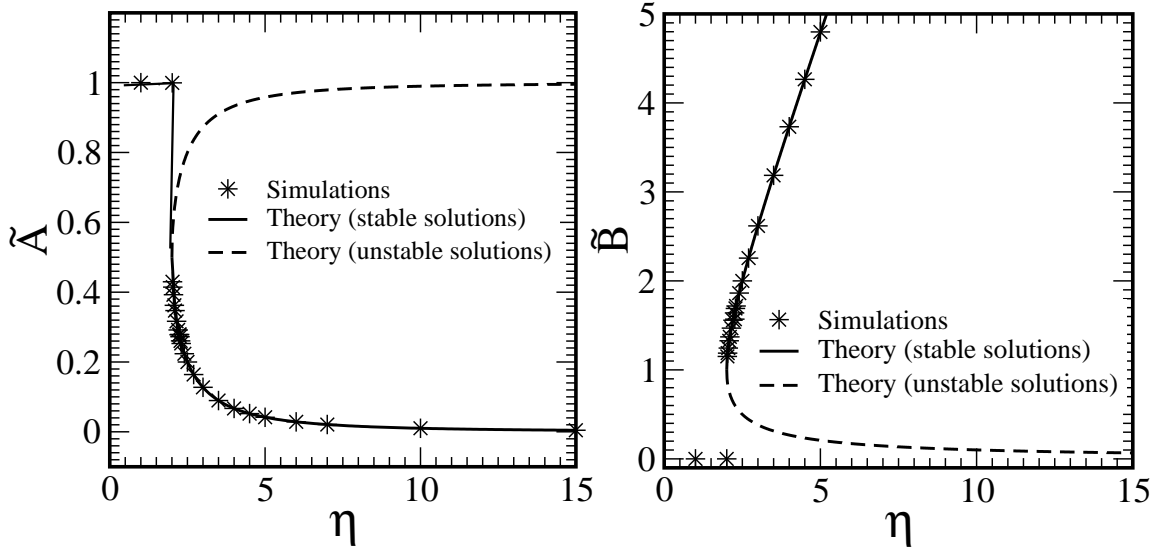}
\caption{ Plot of the stationary homogeneous state solutions of species A and B given by  \myeq{eq:reac_diff19}. Above the bifurcation point ($\eta=2.0$), two solutions exist: one is unstable to homogeneous perturbations (indicated as dashed line) and the other may be stable (plotted as a solid line). At $\eta=2.0$, the stable solution switches to the trivial homogeneous state (1,0) and for $\eta<2.0$ only the trivial state exist. The Lattice Boltzmann simulation (indicated as symbols ) shows good agreement with the theory.}
\label{fig:graph4}
\end{figure}

Figure \ref{fig:graph4} shows a typical bifurcation diagram for the system. We plot the homogeneous steady state solutions obtained in \myeq{eq:reac_diff19}  as a function of the control parameter $\eta$. For $\eta < 2$ there exists only the trivial state (1,0), while when $\eta > 2$ two additional states $(\tilde{A}^{-}_{e},\tilde{B}^{-}_{e})$ and $(\tilde{A}^{+}_{e},\tilde{B}^{+}_{e})$ emerge. The state $(\tilde{A}^{+}_{e},\tilde{B}^{+}_{e})$ is always unstable (indicated by dashed line) while the state $(\tilde{A}^{-}_{e},\tilde{B}^{-}_{e})$ is stable if $\tau< \eta\tilde{B}^-_e$.  In the same figure we have also plotted the steady state solutions obtained from the Lattice Boltzmann simulation of the spatially homogeneous solutions with small homogeneous perturbations at time t = 0. As seen in \myfig{fig:graph4}, the LB simulation well reproduces the analytically predicted stability diagram.  

\label{sec:Turing}
\subsection{Inhomogeneous state and Turing instability}
The Gray-Scott model develops a Turing instability for a range of parameters. In this region of the parameter space the homogeneous steady state solution becomes unstable and a new stationary but inhomogeneous state characterized by the formation of patterns becomes stable. We examine the condition for Turing instability in this system by looking at the growth rate $\alpha$ of an infinitesimal spatially inhomogeneous perturbation to the steady state solutions 
\begin{equation}
  \tilde{A} = \tilde{A}_{e}+\phi_{A}e^{\alpha t}e^{i\tilde{q}x },\quad
  \tilde{B} = \tilde{B}_{e}+\phi_{B}e^{\alpha t}e^{i\tilde{q}x}.
  \label{eq:reac_diff27}
\end{equation}   

As in the case of \myeq{eq:reac_diff22}, $\phi_{A}$ and $\phi_{B}$ are the amplitude of the perturbations to the species $A$ and $B$ respectively, and 
$q$ is the wave number. Again, inserting \myeq{eq:reac_diff27} into the kinetic Eqs.~(\ref{eq:reac_diff16}) and (\ref{eq:reac_diff17}) and after linearizing and slight re-arrangement one arrives at the eigenvalue equation
  \begin{equation}
   \left( \bm{M}-\alpha\bm{I}\right)\bm{\phi} = 0,
   \label{eq:reac_diff28}
  \end{equation}  
where the matrix $\bm{M}$ is written as       
\begin{equation}
 \bm{M} = \begin{bmatrix}
  \tau\left( 2\eta\tilde{A}^{\pm}_{e}\tilde{B}^{\pm}_{e}-1-\dfrac{\tilde{q}^{2}}{\varepsilon^{2}}\right) &\tau\eta\tilde{B}^{\pm^{2}}_{e}  \\\\
  -2\tilde{A}^{\pm}_{e}\tilde{B}^{\pm}_{e}& -(\tilde{q}^{2}+\tilde{B}^{\pm^{2}}_{e}+1)
 \end{bmatrix}.
\label{eq:reac_diff29}
\end{equation}
 For the trivial solution $(A_{e} =1,B_{e}=0)$ the matrix $M$ reduces to 
\begin{equation}
 \bm{M} = \begin{bmatrix}
  -\tau\left(1+\dfrac{\tilde{q}^{2}}{\varepsilon^{2}}\right) &0  \\\\
  0& -(\tilde{q}^{2}+1)
 \end{bmatrix},
\label{eq:reac_diff30}
\end{equation}

and solving the eigenvalue equation in (\ref{eq:reac_diff28}), we obtain the eigenvalues of the trivial state as 
$\alpha_{1} = -\tau(1+\tilde{q}^{2}/\varepsilon^{2})$ and $\alpha_{2} = -(\tilde{q}^{2}+1)$.   Since both  eigenvalues are negative, the trivial homogeneous state $(A_{e}= 1,B_{e}= 0)$ is linearly stable for all system parameters, and independent of the wavelength of the applied perturbation. However, it is important to emphasize that this stability is restricted to infinitesimal perturbations. Indeed, the trivial state is found to be unstable with respect to large amplitude spatially inhomogeneous perturbations. In fact, it is in this regime that the so called self replicating spots are observed. The original parameterization of the Gray-Scott model by Pearson~\cite{Pearson1993} is also based on the numerical simulation of spatially inhomogeneous perturbations of the trivial state. 

  \begin{figure}
\centering
\includegraphics[scale=0.60]{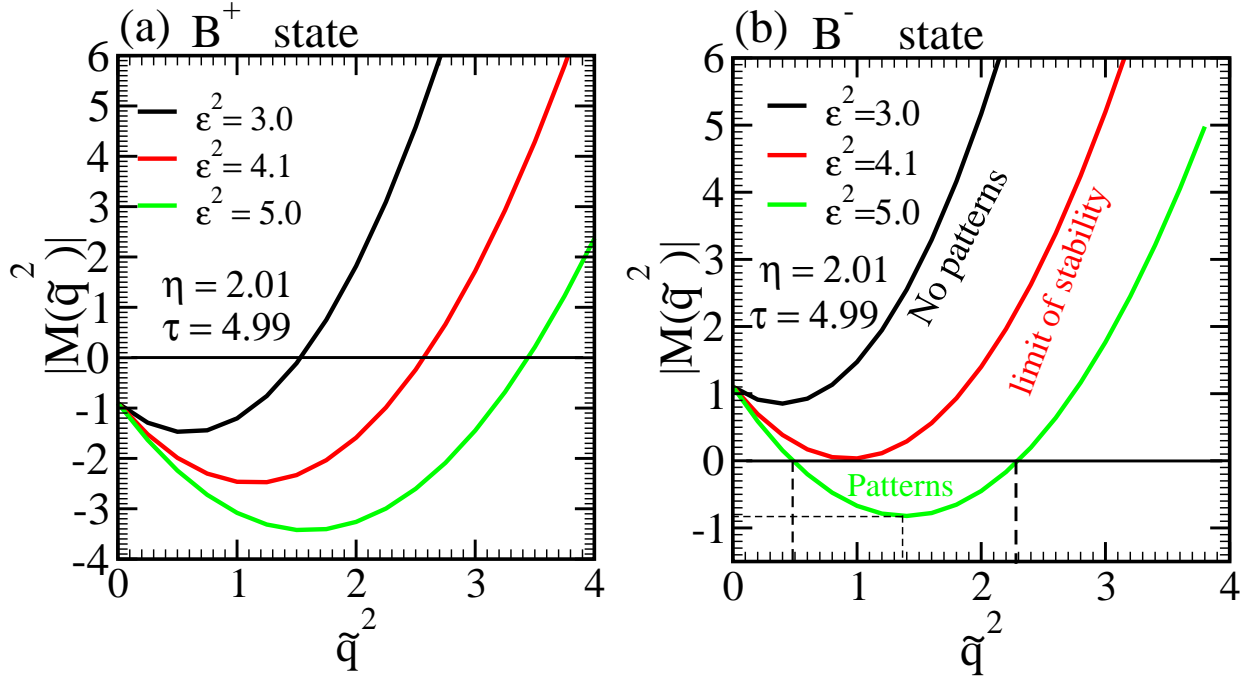}
\caption{The plot of $|\bm{M}(\tilde{q}^{2})| $ versus  $\tilde{q}^{2}$ for states (a)$B^{+}_{e}$  (b)$B^{-}_{e}$. These plots show the range of growth modes $q$ for which the determinant $|\bm{M}(\tilde{q}^{2})|$ is negative with the possibility of pattern formation.}
\label{fig:graph5}
\end{figure}
\begin{figure}
\centering
\includegraphics[scale=0.60]{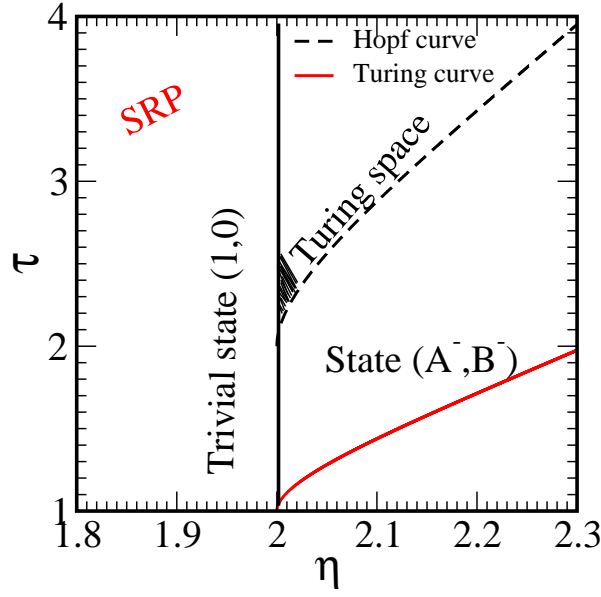}
\caption{The phase diagram of the model at $D_{A}/D_{B} = 2$ showing the Turing curve and the Hopf curve for parameter space spanned by $\tau$ and $\eta$. For $\eta<2$ only the trivial state exists and the self replicating spots (SRP) are observed in this regime only. For $\eta>2$ two additional states $(A^{\pm}_{e}, B^{\pm}_{e})$ emerge. In this figure we have only shown the states $(A^{-}_{e}, B^{-}_{e})$. The majority of the patterns formed from the $B^{-}$ state in our Lattice Boltzmann simulations are observed in the shaded region of the Turing space.}
\label{fig:graph6}
\end{figure}

\begin{figure}
\centering
\includegraphics[scale = 1.3]{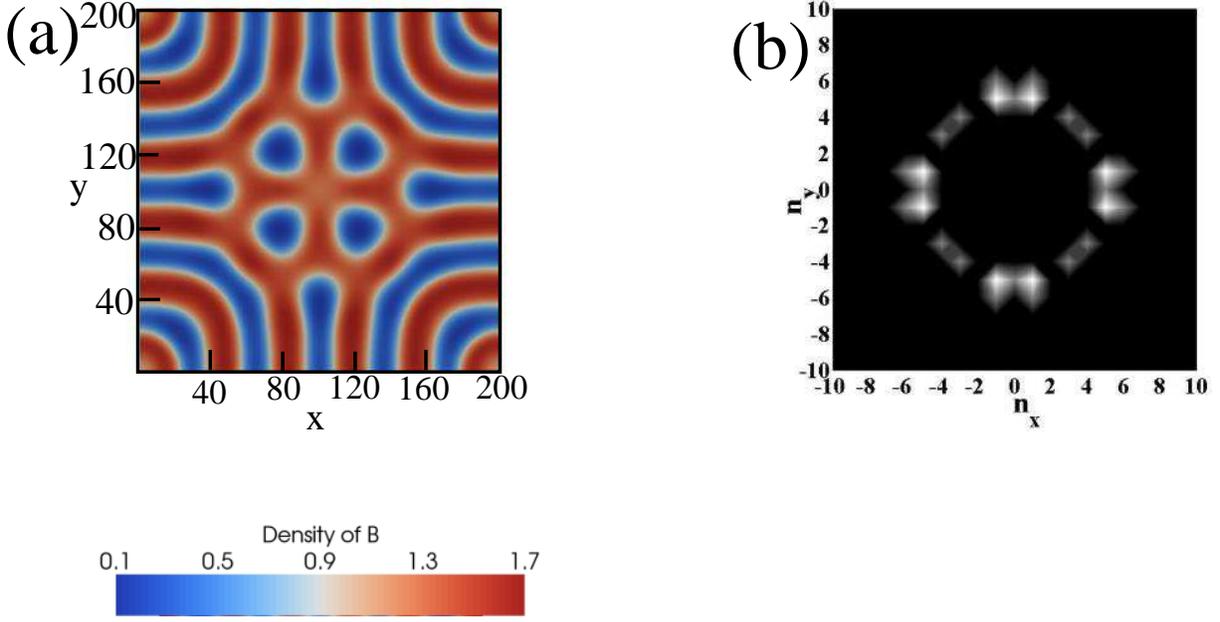}
\caption{(a) Spatial distribution of the density $\tilde{B}$ at time t = 400000, $\eta = 2.0139$ and $\tau = 2.7330$. The formation of stripes can be observed at these parameters.  (b) Amplitude of the Fourier components of density fluctuation ($\tilde{B}$ - $\tilde{B}_{e}$) at time t = 400000,  as function of the wave vector $q = 2\pi\left( (n_{x}/L_{x})^{2}+(n_{y}/L_{y})^{2}\right)^{1/2} $. The dimensionless density $\tilde{B}_{e}$ corresponds to the unstable homogeneous state for the selected set of parameters $\eta$ and $\epsilon$. The white region in the Fourier spectrum corresponds to the excited wave numbers $(n_{x},n_{y})$.}  
\label{fig:graph7}
\end{figure}

\begin{figure}
\centering
\includegraphics[scale = 0.95]{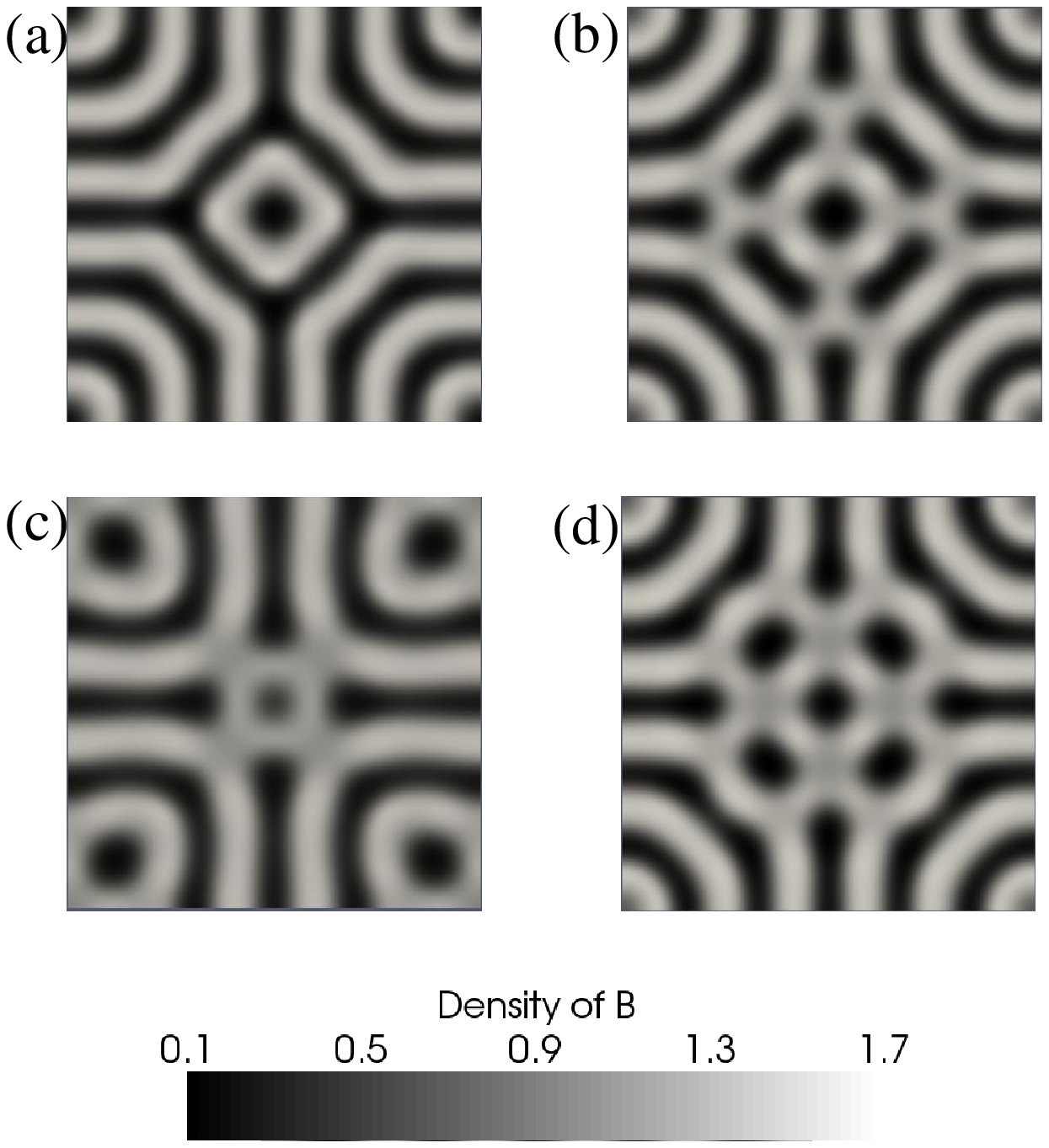}
\caption{Stable time independent Turing structures developed from infinitesimal perturbations to state $(A^{-}_{e}, B^{-}_{e})$ at parameters (a) $\eta = 2.0072819$, $\tau=2.742424$ (b) $\eta = 2.013958$, $\tau=2.73333$ (c) $\eta = 2.017971$, $\tau=2.707462$ and (d) $\eta = 2.0184336$, $\tau=2.7272$. The system size in all cases considered above is 200 x 200 lattice units.}
\label{fig:graph8}
\end{figure}

For the remaining non-trivial solutions, we insert the homogeneous steady state solutions  $\tilde{A}^{\pm}_{e} = 1/(\eta \tilde{B}^{\pm}_e)$ into the matrix $\bm{M}$ in \myeq{eq:reac_diff29} and solve for the eigenvalues of $\bm{M}$ with the characteristic  \myeq{eq:reac_diff25} by replacing $\bm{J}$ with $\bm{M(q)}$. The corresponding eigenvalues are then obtained from \myeq{eq:reac_diff26} by replacing tr$\bm{J}$ and $|\bm{J}|$ with 
   $\text{tr}\bm{M}(q^{2}) = \tau - \eta\tilde{B}^{\pm}_{e}-\tilde{q}^{2}(\tau/\varepsilon^{2} +1)$ and  $|\bm{M}(q^{2})|= \tilde{q}^{4}\tau/\varepsilon^{2}+\tilde{q}^{2}(\tau\eta\tilde{B}^{\pm}_{e}/\varepsilon^{2}-\tau)+\tau(\eta\tilde{B}^{\pm}_{e}-2)$ respectively.

Turing structures or patterns emerge when the system becomes unstable with respect to inhomogeneous perturbations. Again, at least one of the eigenvalues becomes positive (unstable), when  $|\bm{M}(q^{2})| < 0$.  $|\bm{M}(q^{2})|$ is a parabola in $q^{2}$ which attains its minimum value for 

\begin{equation}
q^{2}_{min}=\dfrac{(\varepsilon^{2}-\eta B^{\pm}_{e})}{2}.
\label{eq:reac_diff31}
\end{equation}

Since $q^{2} > 0$,  a minimum in $|\bm{M}(q^{2})|$ exists only if $\varepsilon^{2}>\eta B^{\pm}_{e}$. This is one of the conditions for Turing instability in this system. The boundary of the instability band or range of wave  number $q$ for which $|\bm{M}(q^{2})| < 0$ is given by the roots of the equation $|\bm{M}(q^{2})| = 0$ :

\begin{equation}
  q^{2}_{1,2}=\dfrac{-(\eta\tilde{B}^{\pm}_{e}-\varepsilon^{2})\pm\sqrt{(\eta\tilde{B}^{\pm}_{e}-\varepsilon^{2})^{2}-4\varepsilon^{2}(\eta\tilde{B}^{\pm}_{e}-2)}}{2}.
  \label{eq:reac_diff32}
 \end{equation}
In  \myeq{eq:reac_diff32}, there is an important observation concerning the state  $B^{+}_{e}$. Using $\eta>2$ and \myeq{eq:reac_diff19} one can show that for the state $B^{+}_{e}$, the condition $\eta\tilde{B}^{+}_{e}<2$ holds for all $\eta>2$. The consequence is that \myeq{eq:reac_diff19} has always one negative root and one positive root independent of the value of  $\varepsilon^{2}$. Since $|\bm{M}(q^{2})| < 0$ for $q=0$ (see \myfig{fig:graph5}(a)), this opens already an instability band for pattern formation as regards the state $B^{+}_{e}$. On the other hand, for the state $B^{-}_{e}$,  two distinct positive real roots are necessary for an instability band of patterns. Thus, the following condition has to be satisfied
 \begin{equation}
   \eta\tilde{B}^{-}_{e}<\varepsilon^{2}<-8+11.65\eta\tilde{B}^{-}_{e} \text{~~~~~~~~~~(Turing space).}
    \label{eq:reac_diff33}
  \end{equation}
The first condition, $\varepsilon^{2}>\eta\tilde{B}^{-}_{e}$ as discussed above, is necessary for the formation of Turing patterns while the second one reflects the requirement of a positive discriminant in \myeq{eq:reac_diff32}. Using the definition of $\varepsilon^{2}$
   the first condition can be re-written as 
\begin{equation}
 \dfrac{l^{2}_{A}}{l^{2}_{B}}>\eta\tilde{B}^{-}_{e}.
 \label{eq:reac_diff34}
\end{equation}
This means that the diffusive length scale for the species A ($ l_{A}=\sqrt{D_{A}\tau_{A}}$) must be at least $\eta\tilde{B}^{-}_{e}$ times larger than that of B ($ l_{B}=\sqrt{D_{B}\tau_{B}}$). In other words, for a given value of parameter $\tau$, the diffusion coefficient of species A has to be  $(\eta\tilde{B}^{-}_{e}/\tau)$  times larger than that of B. In $\eta$ and $\tau$ parameter space, the curves $\tau = (D_{B}/D_{A})\eta\tilde{B}^{-}_{e}$, and $\tau = (D_{B}/D_{A})(-8+11.65\eta\tilde{B}^{-}_{e})$ define the limits of  stability with respect to Turing patterns. The first curve $\tau = (D_{B}/D_{A})\eta\tilde{B}^{-}_{e}$ is plotted as the Turing curve in \myfig{fig:graph6} for $D_{A}/D_{B} = 2$. The second curve $\tau = (D_{B}/D_{A})(-8+11.65\eta\tilde{B}^{-}_{e})$ lies above the first curve and falls outside the plotted range. It is therefore not shown in the figure. At zero mode $(q =0)$, another important instability known as the Hopf instability occurs when the real part of a pair of complex eigenvalues passes through zero. In other words, a Hopf instability characterizes the transition from a decaying oscillating mode ($\text{tr}\bm{M}(0)<0$) to an oscillation with growing amplitude ($\text{tr}\bm{M}(0)>0$). Thus, the limit of Hopf instability is given by the condition  $\text{tr}\bm{M}(0)=0$. The dashed black line in \myfig{fig:graph6} indicates the limit of the Hopf instability. The small dashed area in the Turing space is the region where most patterns are expected to be observed.
 
\subsection{Lattice Boltzmann simulation of the spatially inhomogeneous dynamics}
\label{sec:LB_simulations_inhomogeneous}
In this section we perform Lattice Boltzmann simulations of the Gray-Scott model for different values of parameter $\tau$ and $\eta$. Here our simulation and parameterization is based on the non-trivial states $(A^{\pm}_{e}, B^{\pm}_{e})$. Starting from the  homogeneous steady state $(A^{-}_{e}, B^{-}_{e})$, we apply a small amplitude density fluctuations of the form  $\delta \rho = \phi \cos(q_{x}x)\cos(q_{y}y)$ where $\phi$ is the amplitude  and 
$q$ is the wave number of the perturbations. We have chosen $\phi =0.001$,  and  $q_{x}$ = $q_{y}$ = 1 in the simulations. 
\myfig{fig:graph7}(a) shows a developed stable structure from  the small amplitude initial perturbation to the $B^{-}$ state with parameters $\eta =2.014$ and $\tau =2.733 $.  For the purpose of comparison with the prediction of linear stability analysis, we perform the Fourier transform of the pattern in \myfig{fig:graph7}(a)  and calculate the excited wave numbers in the Fourier spectrum  using the relation
\begin{equation}
q = 2\pi\left( (n_{x}/L_{x})^{2}+(n_{y}/L_{y})^{2}\right)^{1/2}, 
 \label{eq:reac_diff35}
\end{equation}

where $n_{x}$ and $n_{y}$ satisfy $-L_{x}/2<n_{x}<L_{x}/2$ and  $-L_{y}/2<n_{y}<L_{y}/2$ respectively.

The Fourier spectrum is shown in gray scale in  \myfig{fig:graph7}(b) and the excited wave numbers are $(n_{x},n_{y})\in\left\lbrace(\pm1,\pm5);(\pm3,\pm4);(\pm4,\pm3);(\pm5,\pm1) \right\rbrace $. Using these values of $n_{x}$ and $n_{y}$ in  \myeq{eq:reac_diff35} to calculate $q$ and the parameters $\eta = 2.0139$ and $\varepsilon^{2} = 5.466$ in \myeq{eq:reac_diff32}, we found that all the excited wave numbers from the simulation fall within the instability band predicted by linear stability analysis in \myeq{eq:reac_diff32}. This provides a further validation of our Lattice Boltzmann simulation with regard to this model.

By performing a  number of similar simulations with different values of $\tau$ and $\eta$ we have found that Turing patterns develop over some part of the  region where the $B^{-}$ state is Hopf or Turing unstable (indicated as the Turing space in \myfig{fig:graph6}). The panels in \myfig{fig:graph8} show  developed stationary structures for typical values of the parameters $\eta$ and $\tau$. These parameter values fall between the saddle node bifurcation curve ($\eta = 2$) and the Turing curve (see \myfig{fig:graph6}). One observation from  \myfig{fig:graph8} is that increasing the value of $\eta$ within the Turing regime leads to the development of a lace-like structure in the patterns. 

\section{Beyond linear stability: Self replicating spots}
\label{sec:scale_inv}
In this section, a further example is provided for the maturity of the Lattice Boltzmann method in studying pattern formation within the Gray-Scott model. The patterns discussed so far are in the part of the phase diagram (\myfig{fig:graph6}), where linear stability analysis predicts that homogeneous solutions are unstable with respect to small perturbations. There are also other types of structures, which occur in a regime, where the trivial homogeneous state is linearly stable. These patterns emerge only if the homogeneous state is  perturbed strongly enough.  A prominent example of this type of structures are the so-called self replicating spots.

\begin{figure}
\centering
\includegraphics[scale = 0.75]{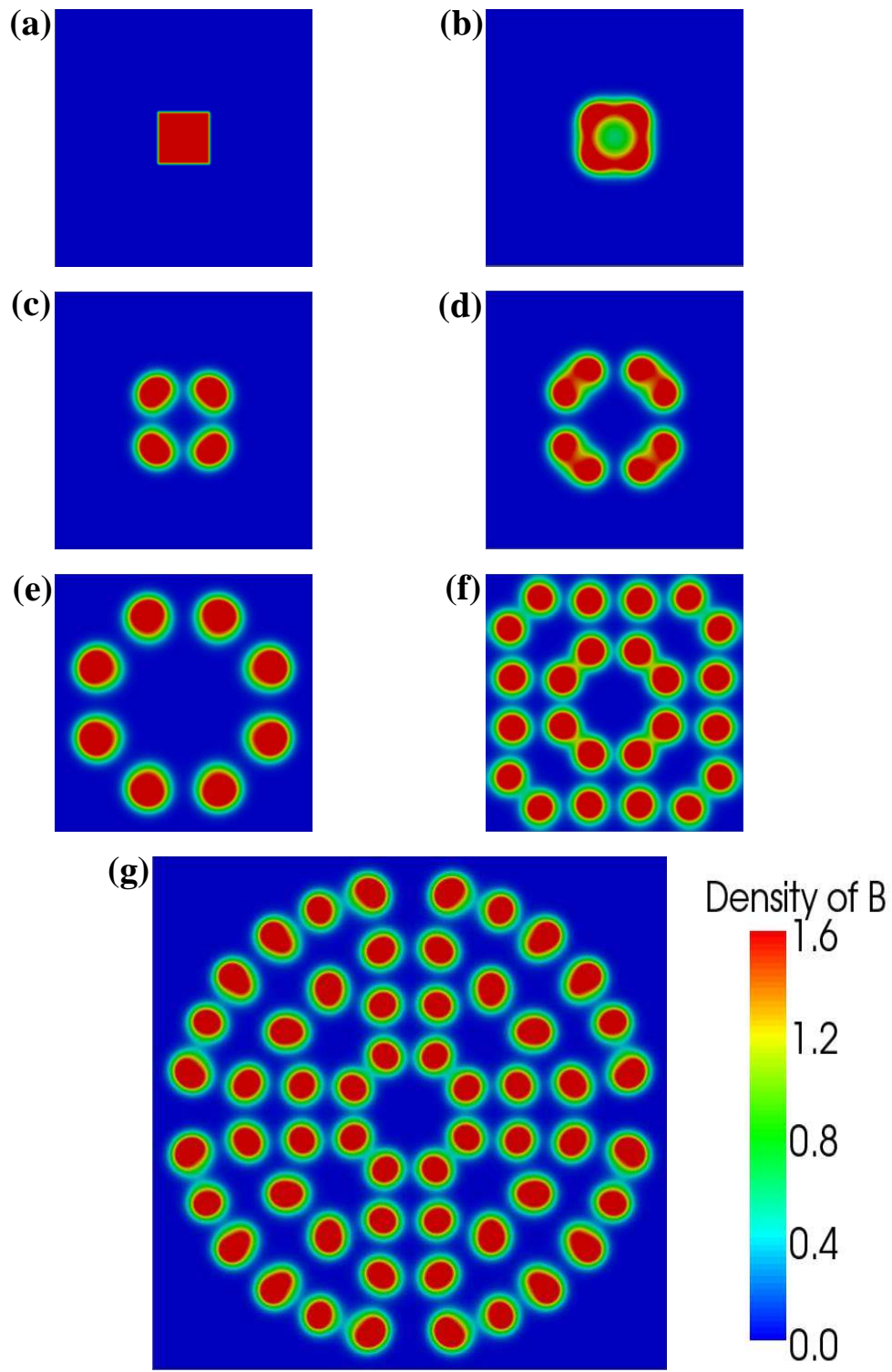}
\caption{Snapshots of the density distribution of species B showing different stages in the self replication process at times (a) t = 0 (b) t = 10000 (c) t = 20000 (d) t = 50000 (e) t = 100000 (f) t = 300000 with size  200 x 200 lattice units and (g) t = 300000 with size 400 x 400 lattice units.}
\label{fig:graph9}
\end{figure}

Figure \ref{fig:graph9} illustrates the patterns emerging from a finite amplitude perturbation of the trivial state. The starting configuration corresponds to a rectangular box of species A and B with densities $\tilde{A}=0.5$, $\tilde{B}=0.25$ placed at the center of a domain filled with species A and B at densities $\tilde{A}=1.0$, $\tilde{B}=0$, respectively. Obviously, such an initial state represents a strong perturbation of the trivial state ($\tilde{A}=1,\; \tilde{B}=0$, i.e.\ $A=A_0,\; B=0$). The sequence of images in \myfig{fig:graph9} demonstrates how spots form, elongate and then replicate as time proceeds. This  self replication process continues until the whole simulation cell is filled with the spots. Interestingly, the number of spots increases with time, while the size of an individual spot seems to remain roughly constant. We have repeated this simulation for a larger box size but otherwise exactly the same parameters. The result of this study is also shown in \myfig{fig:graph9}. As seen from the last image in \myfig{fig:graph9}, the size of a spot does not change, but only the number of spots increases as to fill the entire simulation cell.

In the light of above presented results, one may rise the question, whether it is possible to keep the number of spots constant but tune their size. An answer to this question is obtained by noting that the dynamics of Eqs.~(\ref{eq:reac_diff16}) and (\ref{eq:reac_diff17})  depends only on the dimensionless quantities $\varepsilon$, $\tau$ and $\eta$. In other words, we must check, whether it is possible to tune the length scale of the problem without altering the values of these dimensionless parameters. This would ensure that the thus obtained new solution will have exactly the same shape (and thus the same number of spots) but a different length scale (different size of spots). Indeed, a look at the parameter $\epsilon$ reveals that it is equal to the ratio of two characteristic lengths $l_A$ and $l_B$, 
$\varepsilon=l_{A}/l_{B}$=$\sqrt{\tau_{A}D_{A}/\tau_{B}D_{B}}$. Thus, if we multiply both $l_A$ and  $l_B$ by a constant factor $\lambda$, the parameter $\epsilon$ remains unchanged. Furthermore, in order to keep also the other two parameters $\tau$ and $\eta$ constant, the simplest choice to achieve such a change of length scale is via diffusion coefficient, i.e.\ via $D_A \to \lambda^2 D_A$ and $D_B \to \lambda^2 D_B $.

\begin{figure}
\centering
\includegraphics[scale = 0.95]{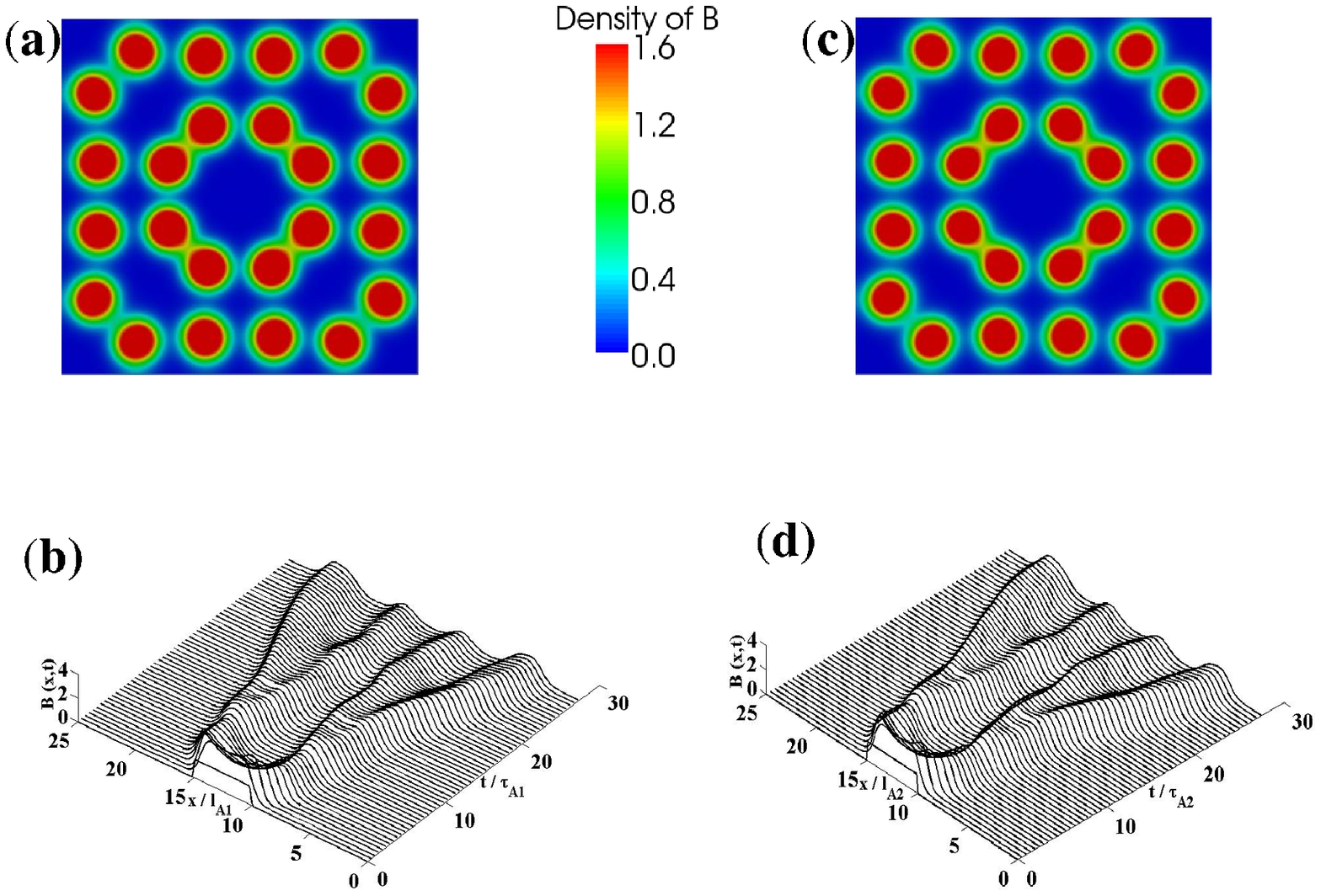}
\caption{Snapshot of the spatial distribution of the density $\tilde{B}$ showing self replicating spots at time 300000, $\eta = 1.86$, $\tau = 3.40$ and $\varepsilon= 2.61$ for the two systems with lattice size (a) 200 x 200 lattice units (c) 400 x 400 lattice units. (b) and (d) show space time plots of the density profile of the self replicating spots along a line in the $y$ direction for the two system size in (a) and (c) respectively.}  
\label{fig:graph10}
\end{figure}

In order to test the above idea, we design two systems such that the system 1 has a linear dimension of $L_1=200$ lattice units with diffusion coefficients $D_{A,1}= 0.016\Delta x^{2}/\Delta t$ and $D_{B,1}= 0.008\Delta x^{2}/\Delta t$. For the system 2, we choose $L_2=400$ lattice units, which means that $\lambda=2$. Following the above arguments, we set the diffusion coefficients of the species A and B in the system 2 to $D_{A,2}= \lambda^2 D_{A,1}= 0.064 \Delta x^{2}/\Delta t$, and $D_{B,2}= \lambda^2 D_{B,1}= 0.032\Delta x^{2}/\Delta t$, respectively. As initial state, we perturb the trivial state exactly in the same way as described in the context of \myfig{fig:graph9} and impose periodic boundary conditions in both the $x$ and $y$ directions. Note that the size of the square perturbation must also be multiplied by $\lambda$ in conformity with the change of length scale. Results of these simulations are shown in \myfig{fig:graph10}(a) and (c). The structure of the patterns is identical for the two systems within numerical discretization errors. The inner core diameter of the spots is found to scale as the diffusion length of species A, $l_{A}$. A more quantitative comparison of the data is provided in \myfig{fig:graph10}(b) and (d), where time evolution of the density profiles is shown for the both studied system sizes in a space-time plot along the $x$ direction.

\begin{figure}
\centering
\includegraphics[scale = 0.95]{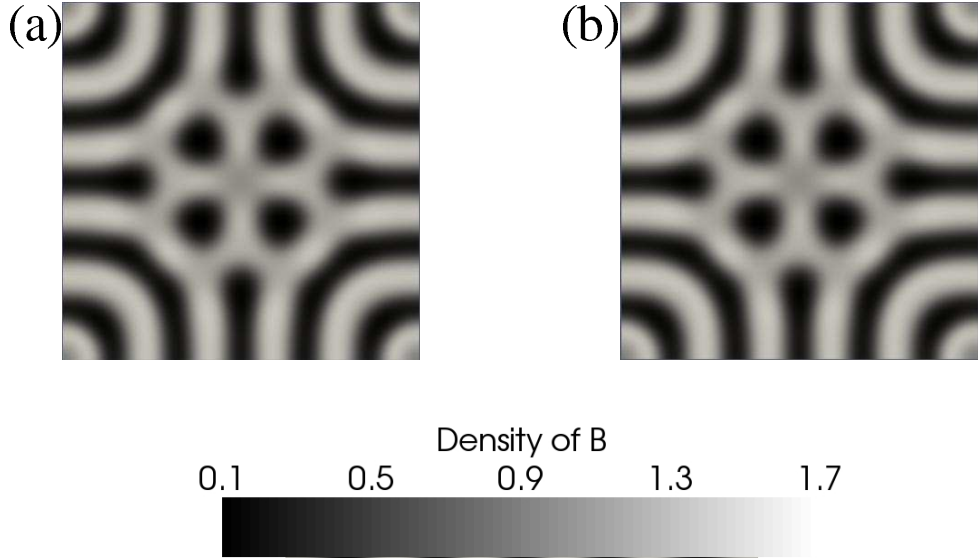}
\caption{Turing pattern showing spatial distribution of the density $\tilde{B}^{-}$ at time t = 400000, $\eta = 2.016933$ and $\tau = 2.73030$ for two systems of size (a)200 x 200 lattice units (b) 400 x 400 lattice units. The patterns are clearly identical in the two cases.}  
\label{fig:graph11}
\end{figure}

\begin{figure}
\centering
\includegraphics[scale = 0.55]{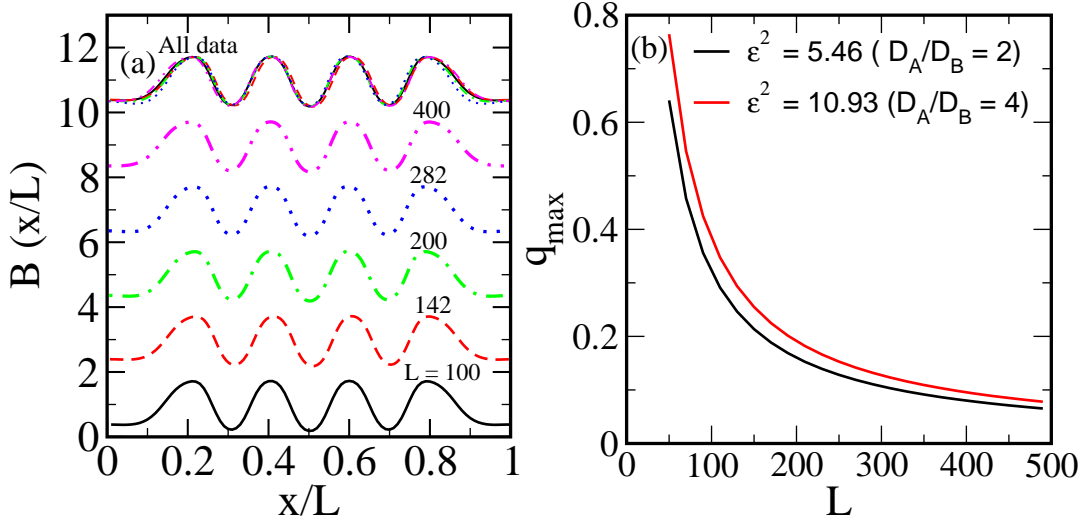}
\caption{(a) Density profile of $\tilde{B}^{-}$ along the line $y/L = 0.85$, obtained from the Turing pattern in  \myfig{fig:graph11} at time t = 400000, with parameters $\eta = 2.014$ and $\tau = 2.733$ parameters . The number of stripes are invariant over an appreciable range of system size. (b) Characteristic wave number of the Turing pattern in \myfig{fig:graph11} plotted against the system size for different values of $\varepsilon^2$ (realized via a variation of $D_{A}/D_{B}$). The curve shows preservation of proportionality between the wave number and the system size.}  
\label{fig:graph12}
\end{figure}

The above arguments on how to tune the length scale while keeping the shape of the patterns unchanged is quite general and applies to any other solution of the Gray-Scott model as well. Here, we provide an example from the Turing regime. This is an interesting test, as linear stability analysis predicts that when the system size is increased, the number of stripes or segments is increased accordingly. However, this applies only if all other parameters are kept constant. Interestingly, by proper regulation of the diffusion coefficient, our numerical simulations in the Turing regime confirm that it is possible to make the wavelength proportional to the system size and keep the number of stripes or segments invariant. Results of these simulations are shown in \myfig{fig:graph11}. For parameters $\eta$ and $\tau$ in the Turing regime, we choose  $\eta=2.014$ and $\tau=2.733$ and consider two systems with a scaling factor $\lambda=2$. The spatial density distribution obtained from the simulations is shown in \myfig{fig:graph11}(a) and (b) for the two systems respectively. Not unexpectedly, the patterns exhibit the same structure with equal number of stripes and segments. To further support this observation, we carried out numerical simulations over a range of system sizes from 50 to 500 lattice units. The thus obtained density profiles along the line $y/L=0.85$ are plotted in \myfig{fig:graph12}(a) for all studied system sizes. For the sake of visibility, each individual curve is shifted by a multiple of 2 along the vertical axis. It is clear from the figure that the number of stripes does not change with the system size. To further emphasize the similarity of the patterns, we directly compare on the same figure all the data using the same shift for all the curves. Clearly, the data collapse into a single curve.

As additional demonstration of wavelength regulation and proportion preservation of the patterns, we perform Fourier transform of the patterns obtained from each system size in the range of 50 to 500 lattice units. We calculate the maximum excited wave number $q_{max}$ in the Fourier spectra of the density field using \myeq{eq:reac_diff35}. \myfig{fig:graph12}(b) shows the plot of the wavenumber excited with the system size for different diffusion coefficients. It is clear from the plot that the maximum excited wavenumber $q_{max}$ decreases in proportion to $L$ and thus the generated pattern is expected to preserve the proportion as observed in our numerical simulation.

\section{Summary}
In this work, we study reaction-diffusion systems via Lattice Boltzmann computer simulations. Starting from the analytical solution of a simple prototypical model (a single species undergoing transformation reaction and diffusion), we perform a systematic study of the Lattice Boltzmann truncation error of the model. We uncover interesting behavior of the truncation error with the system parameters. The error is found to have a minimum at a given value of diffusion coefficient. The position of minimum is shifted for increasing values of the reaction rate constant. These observations are in agreement with the analytical findings from a third-order Chapman-Enskog multiscale expansion. 

A study of the Gray-Scott reaction-diffusion model is also provided. Here, we perform a linear stability analysis of the model and determine the relevant parameter range for pattern formation. Lattice Boltzmann simulations of this interesting reaction-diffusion system are found to be in good agreement with the predictions of the linear stability analysis. In addition to a test of the linear stability phase diagram, Lattice Boltzmann simulations provide valuable information on the details of the patterns formed in different regions of the parameter space. An example is the formation of striped patterns in most parts of the Turing regime above the Hopf bifurcation curve. Another very interesting example is provided by the so called self replicating spots, which lie beyond the linear stability regime. Self replicating spots occur via large amplitude perturbations of the trivial homogeneous solution. 

Furthermore, a survey of the parameters entering the scale invariant form of the Gray-Scott model suggests that the simplest choice to tune the length scale of the obtained patterns (while keeping its shape unchanged) is to multiply all the relevant diffusion coefficients with the same constant factor without any modification of the reaction rates. Interestingly, this lets the time scale of the process unaffected. In other words, in systems with different diffusion coefficients but same reaction rates, patterns exhibit the same shape (but different sizes) exactly for the same physical time. Results obtained via Lattice Boltzmann simulations confirm this behavior. It is noteworthy that this act of regulating diffusion coefficient of species or morphogens as the case may be, is also observed in some biological systems. This observation is by no means limited to this model, the analysis can also be extended to other reaction-diffusion model.

\acknowledgements{This work was supported by the Max-Planck Initiative for Multiscale Materials Modeling of Condensed matter (MMM) and the Interdisciplinary Center for Advanced Material Simulation (ICAMS). We are most grateful to Timm Kr\"uger for useful discussions.}

\appendix
\section{Chapman-Enskog procedure for reaction-diffusion equation}
In this appendix, we derive the macroscopic reaction-diffusion equation from the Lattice Boltzmann model. The LB equation for reaction-diffusion equation is written as
 \begin{equation}
 f_{i,s}(\bmx+\ei\Delta t, \Delta t+t)-f_{i,s}(\bmx, t)=\frac{\fiseq(\bmx, t)-f_{i,s}(\bmx, t)}{\tau_{\text{LB},s}}+\Delta t w_{i}R_{s}
\label{eq:app1}
\end{equation}
To obtain a corresponding  macroscopic partial differential equation from the finite difference \myeq{eq:app1},  we perform a Taylor series expansion of the left hand side of \myeq{eq:app1} and obtain
\begin{equation}
 \sum_{n=1}^\infty \frac{\Delta t^n}{n!}(\partial_{t}+\eia\partial_{x_{\alpha}})^nf_{i}(\bmx,t)= \frac{\fiseq(\bmx, t)-f_{i,s}(\bmx, t)}{\tau_{\text{LB},s}}+\Delta tw_{i}R_{s}.
\label{eq:app3}
\end{equation}
The Chapman Enskog procedure introduces two time scales, a fast time scale, $t_{1}$, associated with convective transport and a slow time scale, $t_{2}$, associated with diffusion. The time derivative is then expanded as
\begin{equation}
 \partial_{t} = \epsilon\partial^{(1)}_{t}+\epsilon^{2}\partial^{(2)}_{t},
\label{eq:app4}
\end{equation}
The spatial derivative is written as
\begin{equation}
\partial_{x_{\alpha}} = \epsilon\partial^{(1)}_{x_{\alpha}}.
\label{eq:app5}
\end{equation} 
The equilibrium distribution as reaction term are expanded as
\begin{eqnarray}
 f_{i,s}&=&f^{(0)}_{i,s}+ \epsilon f^{(1)}_{i,s}+ \epsilon^2 f^{(2)}_{i,s}+ \epsilon^3 f^{(3)}_{i,s}+{\cal{O}}(\epsilon^4)
\label{eq:app6}\\
 R_{s} &=& R^{(0)}_{s}+\epsilon R^{(1)}_{s}+\epsilon^{2} R^{(2)}_{s}+\epsilon^{3} R^{(3)}_{s}+ {\cal{O}}(\epsilon^4)
\label{eq:app7}
\end{eqnarray}
Inserting Eqs.~(\ref{eq:app7}), (\ref{eq:app6}), (\ref{eq:app5}) and  (\ref{eq:app4}) in \myeq{eq:app3}, one obtains

\begin{multline}
 \left[ \Delta t(\epsilon\partial^{(1)}_{t}+\epsilon^{2}\partial^{(2)}_{t}+\epsilon\eia\partial^{(1)}_{x_{\alpha}})+\frac{\Delta t^{2}}{2}(\epsilon^{2}\partial^{(1)}_{t}\partial^{(1)}_{t}+2\epsilon^{2}\eia\partial^{(1)}_{t}\partial^{(1)}_{x_{\alpha}}+\epsilon^{2}\eia\eib\partial^{(1)}_{x_{\alpha}}\partial^{(1)}_{x_{\beta}}+2\epsilon^{3}\eia\partial^{(2)}_{t}\partial^{(1)}_{x_{\alpha}}\right.\\ +\left.2\epsilon^{3}\partial^{(2)}_{t}\partial^{(1)}_{t}
+2\epsilon^{4}\partial^{(2)}_{t}\partial^{(2)}_{t}\right] (f^{(0)}_{i,s}+ \epsilon f^{(1)}_{i,s}+ \epsilon^2 f^{(2)}_{i,s}+{\cal{O}}(\epsilon^3))\\=\frac{1}{\tau_s}\left(\fiseq(\bmx, t)-(f^{(0)}_{i,s}+ \epsilon f^{(1)}_{i,s}+\epsilon^2 f^{(2)}_{i,s}+ \epsilon^3 f^{(3)}_{i,s}+{\cal{O}}(\epsilon^4))\right)+\Delta tw_{i}(\epsilon R^{(1)}_{s}+\epsilon^{2} R^{(2)}_{s}+\epsilon^{3} R^{(3)}_{s}+{\cal{O}}(\epsilon^4)).
\label{eq:app_exp}
\end{multline}
  
Grouping terms of the same order in $\epsilon$ yields the following successive approximations.

\begin{equation}
 O(\epsilon^{(0)}): \quad\quad\quad\quad\quad\qquad\quad\quad\quad\quad \quad\quad f^{(0)}_{i} = \fiseq, \quad\quad \text{implying that} \quad\quad  R^{(0)}_{s} = 0. \quad\quad\quad\quad\quad\quad 
\label{eq:app_7b}
\end{equation}
Note that this condition follows directly from the conservation of mass.
\begin{multline}
O(\epsilon^{1}):\quad\qquad\quad\quad\quad\quad\quad\quad
\Delta t\left(\partial^{(1)}_{t}+ \eia\partial^{(1)}_{x_{\alpha}}\right) f^{(0)}_{i,s}=-\frac{1}{\tau_{\text{LB},s}}f^{(1)}_{i,s} + \Delta tw_{i}R^{(1)}_{s}\quad\quad\quad
\label{eq:app8}
\end{multline}
On the level of $\epsilon$, there is no mass diffusion, diffusion process takes place on the scale of $\epsilon^{2}$. Furthermore, for diffusion driven reactions, the diffusive flux must bring the species together before reaction and reaction becomes a second order effect. Since on the scale of $\epsilon$, there is no mass diffusion, that implies $R^{(1)}_{s}= 0$.
We would consider this case in this derivation and \myeq{eq:app8} becomes

\begin{multline}
O(\epsilon^{1}):\quad\qquad\quad\quad\quad\quad\quad\quad
\Delta t\left(\partial^{(1)}_{t}+ \eia\partial^{(1)}_{x_{\alpha}}\right) f^{(0)}_{i,s}=-\frac{1}{\tau_{\text{LB},s}}f^{(1)}_{i,s}. \quad\quad\quad
\label{eq:app8_new}
\end{multline}
\begin{multline}
O(\epsilon^{2}): \quad\quad
\Delta t\left( \partial^{(2)}_{t}f^{(0)}_{i,s}+ \left( \partial^{(1)}_{t}+\eia\partial^{(1)}_{x_{\alpha}}\right) f^{(1)}_{i,s}\right) + \\
\frac{\Delta t^{2}}{2}\left( \partial^{(1)^{2}}_{t} +2\eia\partial^{(1)}_{t}\partial^{(1)}_{x_{\alpha}}+\eia\eib\partial^{(1)}_{x_{\alpha}}\partial^{(1)}_{x_{\beta}}\right) f^{(0)}_{i,s}= -\frac{1}{\tau_{\text{LB},s}}f^{(2)}_{i}+\Delta tw_{i}R^{(2)}_{s}.
\label{eq:app9}
\end{multline}
\begin{multline}
O(\epsilon^{3}): \quad\quad\quad
\Delta t \left(\partial^{(3)}_{t}f^{(0)}_{i,s} +\partial^{(2)}_{t}f^{(1)}_{i,s}+\left( \partial^{(1)}_{t}+\eia\partial^{(1)}_{x_{\alpha}}\right) f^{(2)}_{i,s}\right)\quad\quad\quad\quad\\ + \frac{\Delta t^{2}}{2}\left( \partial^{(1)^{1}}_{t} +2\eia\partial^{(1)}_{t}\partial^{(1)}_{x_{\alpha}}+\eia\eib\partial^{(1)}_{x_{\alpha}}\partial^{(1)}_{x_{\beta}}\right) f^{(1)}_{i,s} \\
+\Delta t^{2} \partial^{(2)}_{t}\left( \partial^{(1)}_{t}+\eia\partial^{(1)}_{x_{\alpha}}\right)f^{(0)}_{i,s}+ \frac{\Delta t^{3}}{6}\left(\partial^{(1)}_{t}+\eia\partial^{(1)}_{x_{\alpha}} \right) ^{3} f^{(0)}_{i,s}\quad =-\frac{1}{\tau_{\text{LB},s}}f^{(3)}_{i,s}+\Delta tw_{i}R^{(3)}_{s}.
\label{eq:app10}
\end{multline}
Putting the expression for $f^{(1)}_{i,s}$ from \myeq{eq:app8_new} into  \myeq{eq:app9} yields
\begin{equation}
\frac{1}{\tau_{\text{LB},s}}f^{(2)}_{i}=-\Delta t\partial^{(2)}_{t}f^{(0)}_{i,s}+\Delta t^{2}\left( \tau_{\text{LB},s}-\frac{1}{2}\right) \left(\partial^{(1)}_{t}+ \eia\partial^{(1)}_{x_{\alpha}}\right)^{2} f^{(0)}_{i,s}+\Delta tw_{i}R^{(2)}_{s}.
\label{eq:app10a}
\end{equation}
In \myeq{eq:app10}, we insert the expression for  $f^{(1)}_{i,s}$ and $f^{(2)}_{i,s}$ from  Eqs.~(\ref{eq:app8_new}) and  (\ref{eq:app9}) and obtain
\begin{multline}
 \frac{1}{\tau_{\text{LB},s}}f^{(3)}_{i,s}=-\Delta t\partial^{(3)}_{t}f^{(0)}_{i,s} +\Delta t^{2}(2\tau_{\text{LB},s}-1)\left(\partial^{(1)}_{t}+ \eia\partial^{(1)}_{x_{\alpha}}\right)\partial^{(2)}_{t}f^{(0)}_{i,s}\\ -\Delta t^{3}\left( \tau_{\text{LB},s}^{2}-\tau_{\text{LB},s}+\frac{1}{6}\right) \left(\partial^{(1)}_{t}+ \eia\partial^{(1)}_{x_{\alpha}}\right)^{3} f^{(0)}_{i,s} -\tau_{\text{LB},s}\Delta t^{2}\left(\partial^{(1)}_{t}+ \eia\partial^{(1)}_{x_{\alpha}}\right)w_{i}R^{(2)}_{s}+\Delta tw_{i}R^{(3)}_{s}.
\label{eq:app10b}
\end{multline}

Next we take the moments of the distribution functions in  Eqs.~(\ref{eq:app8}), (\ref{eq:app10a}) and (\ref{eq:app10b}). Note that in order to preserve the isotropy of the lattice tensors, the chosen lattice speeds and weights in the equilibrium distribution function must obey the following moments or symmetry conditions. 
\begin{eqnarray}
 (a)\sum_{i}w_{i} & = & 1 \nonumber \\
 (b)\sum_{i} w_{i}\eia & = & 0 \nonumber\\ 
 (c)\sum_{i}w_{i}\eia\eib & = & c^{2}_{s}\delta_{\alpha\beta}\nonumber\\
 (d)\sum_{i}w_{i}\eia\eib\eig & = & 0\nonumber\\
 (e)\sum_{i}w_{i}\eia\eib\eig \eide& = & c^{4}_{s}(\delta_{\alpha\beta}\delta_{\gamma\delta}+
  \delta_{\alpha\gamma}\delta_{\beta\delta}+\delta_{\alpha\delta}\delta_{\beta\gamma})\nonumber\\
\label{eq:app11}
\end{eqnarray}
Using \myeq{eq:app11} and given that the local equilibrium takes the form $f^{eq}_{i}= f^{(0)}_{i} = w_{i}\rho_{s}$, we impose the following  conditions of conservation of mass on the equilibrium distribution function

\begin{equation}
 \sum_{i}f^{(0)}_{i,s} = \rho_{s}, \quad  \sum_{i}\eia f^{(0)}_{i,s} = 0, \quad \sum_{i}\eia\eib f^{(0)}_{i,s}=\rho_{s}c^{2}_{s}\delta_{\alpha\beta}
\label{eq:app12}
\end{equation}

We further assume that higher order corrections of the equilibrium distribution do not contribute to the local values of the mass, whereby obtaining 
\begin{equation}
 \sum_{i} f^{(n)}_{i}=0, \quad \text{for} \quad n \geq 1.
 \label{eq:app13}
\end{equation}

Taking  $\sum_{i}$  of  \myeq{eq:app8} and using Eqs.~(\ref{eq:app11}), (\ref{eq:app12}) and (\ref{eq:app13})  yields  

\begin{equation}
\partial^{(1)}_{t} \rho_{s}  = 0,
\label{eq:app14}
\end{equation}

Taking $\sum_{i}$  \myeq{eq:app10a} and using Eqs.~(\ref{eq:app11}), (\ref{eq:app12}), (\ref{eq:app13}) and (\ref{eq:app14}) leads to 

\begin{equation}
 \partial^{(2)}_{t}\rho_{s}=\Delta tc^{2}_{s} \left( \tau_{\text{LB},s}-\frac{1}{2}\right) \partial^{(1)}_{x_{\alpha}} \partial^{(1)}_{x_{\beta}}\rho_{s}\delta_{\alpha\beta}+ R^{(2)}_{s}
\label{eq:app17}
\end{equation}

Taking $\sum_{i}$ of  \myeq{eq:app10b} and using Eqs.~(\ref{eq:app11}), (\ref{eq:app12}), (\ref{eq:app13}), 
and (\ref{eq:app14}) leads to
\begin{equation}
\partial^{(3)}_{t}\rho_{s} =-3c^{2}_{s}\Delta t^{2}\left( \tau_{\text{LB},s}^{2}-\tau_{\text{LB},s}+\frac{1}{6}\right)\partial^{(1)}_{t}\partial^{(1)}_{x_{\alpha}}\partial^{(1)}_{x_{\beta}}\rho_{s}\delta_{\alpha\beta}-\tau_{\text{LB},s}\Delta t^{2}\partial^{(1)}_{t}R^{(2)}_{s} + R^{(3)}_{s}. 
\label{eq:app19}
\end{equation}

We multiply \myeq{eq:app14} by  $\epsilon$, \myeq{eq:app17} by $\epsilon^{2}$ and \myeq{eq:app19} by  $\epsilon^{3}$  and add all this together, thus arriving at
\begin{equation}
\partial_{t}\rho_{s}=c^{2}_{s}\Delta t\left( \tau_{\text{LB},s}-\frac{1}{2}\right) \partial^{2}_{x_{\alpha}} \rho_{s}
+ R_{s}- 3\Delta t^{2}c^{2}_{s}\left( \tau_{\text{LB},s}^{2}-\tau_{\text{LB},s}+\frac{1}{6}\right)\partial_{t}\partial^{2}_{x_{\alpha}}\rho_{s}-\tau_{\text{LB},s}\Delta t^{2}\partial_{t}R_{s}.
\label{eq:app21}
\end{equation}

We can further re-write  \myeq{eq:app21} as  the macroscopic reaction-diffusion equation and a third-order truncation error term $E$
\begin{equation}
\partial_{t}\rho_{s}=D_{s} \partial^{2}_{x_{\alpha}} \rho_{s}
+ R_{s}- E
\label{eq:app22}
\end{equation}
where the diffusion coefficient is given by $D_{s}=c^{2}_{s}\Delta t\left( \tau_{\text{LB},s}-0.5\right)$ and the error term takes the form \begin{equation}
 E =  3\Delta t^{2}c^{2}_{s}\left( \tau_{\text{LB},s}^{2}-\tau_{\text{LB},s}+\frac{1}{6}\right)\partial_{t}\partial^{2}_{x_{\alpha}}\rho_{s}+\tau_{\text{LB},s}\Delta t^{2}\partial_{t}R_{s}+ {\cal{O}}(\epsilon^{4}).
\label{eq:app23}
\end{equation}

\end{document}